\documentclass[pre,aps,twocolumn,superscriptaddress]{revtex4-1}
\pdfoutput=1
\usepackage{amssymb}
\usepackage{epsfig}
\usepackage[export]{adjustbox}
\usepackage{amsmath}
\usepackage{caption}
\usepackage{subcaption}
\usepackage{graphicx}
\usepackage{textcomp}
\usepackage{url}
\usepackage{float}
\usepackage{color}
\usepackage{cases}
\usepackage[normalem]{ulem}

\usepackage[titletoc,title]{appendix}

\usepackage[colorlinks=true, urlcolor=blue, anchorcolor=blue, citecolor=blue,filecolor=blue,linkcolor=blue,menucolor=blue
]{hyperref}

\usepackage{color}

\begin{document}
\title{Toward the full short-time statistics of an active Brownian particle on the plane}

\author{Satya N. Majumdar}
\affiliation{LPTMS,  CNRS,  Univ.   Paris-Sud,  Universit\'{e}  Paris-Saclay,  91405  Orsay,  France}
\email{satya.majumdar@u-psud.fr}
\author{Baruch Meerson}
\affiliation{Racah Institute of Physics, Hebrew University of
Jerusalem, Jerusalem 91904, Israel}
\email{meerson@mail.huji.ac.il}

\begin{abstract}
We study the position distribution of a single active Brownian particle (ABP) on the plane. We show that this distribution has a compact support, the boundary of which is an expanding circle. We focus on a short-time regime and  employ the optimal fluctuation method (OFM) to study large deviations of the particle position coordinates $x$ and $y$. We determine the optimal paths of the ABP, conditioned on reaching specified values of $x$ and $y$, and the large deviation functions of the marginal distributions of $x$, and of $y$. These marginal distributions match continuously with  ``near tails" of the $x$ and $y$ distributions of typical fluctuations, studied earlier.  We also calculate the large deviation function of the joint $x$ and $y$ distribution $P(x,y,t)$ in a vicinity of a special ``zero-noise" point, and show that $\ln P(x,y,t)$ has a nontrivial self-similar structure as a function of $x$, $y$ and $t$.  The joint distribution vanishes extremely fast at the expanding circle,  exhibiting an essential singularity there.
This singularity is inherited by the marginal $x$- and $y$-distributions. We argue that this fingerprint of the short-time dynamics remains there at all times.
\end{abstract}

\maketitle
\nopagebreak

\section{Introduction}
\label{intro}

Spatio-temporal dynamics of active self-propelled particles has been a subject of much current interest, both
theoretically and experimentally. Contrary to
passive Brownian  particles, that are driven by thermal
noise, generated via the collisions with molecules in a surrounding fluid, a self-propelled active particle
consumes energy directly from the environment and moves ballistically in a stochastically evolving direction
with an intrinsic
velocity $v_0$. Such active particles occur in many biological and soft matter systems including E. coli bacteria,
fish schools, colloidal surfers amongst others (for reviews, see \textit{e.g.}
Refs.~\cite{Romanczuk,soft,BechingerRev,Marchetti2017}). Interaction between self-propelled particles leads to new
phases and other interesting collective properties. Numerous recent studies have revealed that, even
without interaction, just a single self-propelled particle displays interesting and complex
spatio-temporal patterns that are yet to be fully understood. Three simple models
of a single self-propelled particle have been studied extensively: active Brownian motion (ABM),
run-and-tumble particle and active Ornstein-Uhlenbeck process. In this paper, we focus
on the ABM only and study its position distribution in two dimensions at short times (or equivalently
in the strongly-active limit) when
the position distribution is highly anisotropic and non-Gaussian.  We show that,
at short times, the large deviation properties of the position distribution can be extracted analytically
using the optimal fluctuation method.

The ABM in two dimensions is defined as follows. Let $(x(t),y(t))$ denote the position
of the self-propelled particle at time $t$. There is an intrinsic velocity vector attached
to the particle (like a ``spin") $\vec v= (v_0\, \cos \phi(t), v_0\, \sin \phi(t))$, where $\phi(t)$ denotes
the angle of orientation  of the spin vector at time $t$. The orientation angle $\phi(t)$ performs
a rotational Brownian motion with diffusion constant $D$. We assume that there is no
direct translational
noise present. The three degrees of freedom $x$, $y$ and $\phi$ then evolve in time according to the equations
\begin{eqnarray}
\label{ABM_model}
\dot{x} &=& v_0 \cos \phi(t), \label{xdoteq} \\
\dot{y} &=& v_0 \sin \phi(t) , \label{ydoteq} \\
\dot{\phi}&=& \sqrt{2\,D} \, \eta(t) \label{phidot} \, ,
\end{eqnarray}
where $\eta(t)$ is a Gaussian white noise with zero mean and a correlator
$\langle \eta(t) \eta(t')\rangle = \delta(t-t')$. We assume
that the particle starts from the origin, $x(0)=0$ and $y(0)=0$, with an initial
orientation angle $\phi(0)$. The most basic question in this problem
concerns the position distribution $P(x,y,t)$ of the particle in the
$xy$ plane at time $t$~\cite{BechingerRev,Marchetti2017,Solon2015,GPB2015,Seifert,Franosch,Basu2018,GL2018,Basu2019,SDC2020}.

To establish some basic properties of the distribution  $P(x,y,t)$,
let us first integrate \eqref{xdoteq} and \eqref{ydoteq} up to time $t$ starting from the origin
\begin{eqnarray}
x(t) & = & v_0\, \int_0^t \cos \phi(t')\, dt'\, , \label{xt.1} \\
y(t) & =& v_0\, \int_0^t  \sin \phi(t')\, dt' \, . \label{yt.1}
\end{eqnarray}
Taking squares of Eqs.~(\ref{xt.1}) and~(\ref{yt.1}) and adding the two equations, we obtain, for arbitrary $t\ge 0$:
\begin{eqnarray}
\label{square.1}
x^2(t)+ y^2(t) &=& v_0^2 \, \int_0^t\int_0^t \cos\left [\phi(t_1)-\phi(t_2)\right]\, dt_1\, dt_2 \nonumber \\
  &\le& v_0^2\, t^2\, .
\end{eqnarray}
That is, the particle cannot be found beyond the expanding circle of radius $v_0 t$, as to be expected on physical grounds.  Moreover, the \emph{equality} in Eq.~\eqref{square.1} is satisfied
if and only if $\phi(t_1)=\phi(t_2)$ for all $t_1$ and $t_2$ in the time interval $[0,t]$,\textit{ i.e.},
if $\phi(t')=\phi(0)$ for all $t'\in [0,t]$. Therefore, the equality $x^2(t)+y^2(t)= v_0^2\, t^2$
holds only at the zero-noise point $(v_0\,t\, \cos \phi(0), v_0\, t\, \sin \phi(0))$ on the circumference of the
circle of radius $v_0 t$. At any other point on the circumference, the strict inequality holds: again, as to be expected.
That is, the distribution $P(x,y,t)$ must be supported over the circle of radius $v_0\, t$ and, moreover,
it must strictly vanish at all points on the circumference of the circle except at
$(v_0\,t\, \cos \phi(0), v_0\, t\, \sin \phi(0))$. However, calculating $P(x,y,t)$ inside
this circle is a challenging problem.

In fact, the ABM model was originally introduced in the mathematics literature,
as the `random curvature' model, by Mumford~\cite{Mumford}. He studied
random algebraic curves in two dimensions in the context of computer vision, much before
the ABM re-surfaced in the literature on active systems. The position distribution $P(x,y,t)$ can, in principle, be
obtained by integrating out the orientational degree of freedom from
the full probability density, $P(x,y,t)= \int P(x,y,\phi,t)\, d\phi$, where
$P(x,y,\phi,t)$ satisfies the Fokker-Planck equation
\begin{equation}\label{eq:FokkerPlanck}
\frac{\partial}{\partial t} P(x,y,\phi,t) = -
v_0 \left( \cos \phi \frac{\partial P}{\partial x}+
\sin \phi \frac{\partial P}{\partial y}\right)+
D\, \frac{\partial^2  P}{\partial \phi^2}
\end{equation}
with the initial condition,
$$
P(x,y,\phi, t=0)= \delta(x)\, \delta(y)\, \delta(\phi-\phi(0))\,.
$$
However, solving this Fokker-Planck equation and extracting $P(x,y,t)$ explicitly is
hard (to quote Mumford, ``I have looked for an
explicit formula for $P$ but in vain'' \cite{Mumford}). When averaged over the initial
orientation $\phi(0)$ distributed uniformly in $[0,2\pi]$ (that makes $P(x,y,t)$
isotropic at all times), it is possible to write a formal
expression for the Fourier transform $f(k,t)= \langle e^{i \vec k\cdot \vec r(t)}\rangle$ in
terms of the eigenfunctions of Mathieu operator~\cite{Franosch}. Nevertheless, to explicitly extract from this formal
expression
the position distribution $P(x,y,t)$ in the real space  is not easy.

The origin of the difficulty in the deceptively simple process \eqref{xdoteq}-\eqref{phidot}
is in the nonlinearity of the functions $\cos \phi$ and $\sin \phi$, leading to a strongly non-Gaussian character of the orientationally driven noises acting on $x$ and $y$. Moreover, writing $\dot x= v_0\, \cos \phi(t)=
\xi_x(t)$ and $\dot y= v_0\, \sin \phi(t)= \xi_y(t)$, we see that the noises
$\xi_x(t)$ and $\xi_y(t)$ are both correlated in time.
Using the Brownian properties of $\phi(t)$, the two-time correlation function of $\xi_x(t)$ can be easily computed
~\cite{Basu2018}
\begin{equation}
\label{eq:autocrr}
\langle \xi_x(t_1) \xi_x(t_2) \rangle \simeq \frac{v_0^2}{2} \exp \left[ - D\, |t_1-t_2| \right]\, ,
\end{equation}
for large $t_1$ and $t_2$ with $|t_1-t_2|$ fixed, and similarly for $\xi_y(t)$. Thus, for short times,
$t\ll D^{-1}$, the noise $\xi_x(t)$  is strongly correlated in time (as opposed to passive
Brownian motion where the noise is delta-correlated). This makes the process non-Markovian and
hence more difficult to analyse. The short-time limit $t\ll D^{-1}$ is called the strongly active limit.

In contrast, for $t\gg D^{-1}$, the noise correlator converges
to $\langle \xi_x(t_1) \xi_x(t_2) \rangle \to 2\, D_{\text{eff}}\, \delta(t_1 -t_2)$
with an effective diffusion constant $D_{\text{eff}} = v_0^2/(2 D_R)$. Thus, in this
strongly passive limit $t\gg D^{-1}$,
the ABM effectively reduces to an ordinary $2$-d Brownian motion, and the position distribution,
irrespective of the initial orientation $\phi(0)$, converges to an isotropic Gaussian form
\begin{equation}
P(x,y,t) \simeq \frac{1}{4\, D_{\rm eff}\, t}\, \exp\left[-\frac{ (x^2+y^2)}{4\, D_{\rm eff}\, t}\right]\, ,
\label{Gauss.1}
\end{equation}
which holds for typical fluctuations, \textit{i.e.}, when $x\sim \sqrt{t}$ and $y\sim \sqrt{t}$. For large
atypical fluctuations, when $r=\sqrt{x^2+y^2}\sim v_0\, t$, the position distribution is
described by the large deviation form
\begin{equation}
\label{e:LDxy}
P(x,y,t) \sim \exp{\left[- t D_R  \,\Psi \left(\frac {\sqrt{x^2+y^2}} {v_0\,t}\right) \right]} \, ,
\end{equation}
where the rate function $\Psi(z)$ was recently computed analytically~\cite{Basu2019}.

It was pointed out in Ref.~\cite{Basu2018} that in the strongly active short-time regime $t\ll D^{-1}$,
the position distribution at time $t$ remembers the initial orientation $\phi(0)$ and is
strongly anisotropic and non-Gaussian, which are fingerprints of `activity' that gets washed out at late times
when $t\gg D^{-1}$. For a fixed initial orientation, say $\phi(0)=0$, it was shown in Ref~\cite{Basu2018}
that $P(x,y,t)$, at $t\ll D^{-1}$, is peaked at $(x=v_0\, t, y=0)$
with typical
fluctuations of $x$ and $y$ scaling as $v_0\, t-x\sim t^2$ and $y\sim t^{3/2}$. The position
distribution, in this typical regime, is described by the scaling form~\cite{Basu2018}
\begin{equation}
P(x,y,t) \simeq \frac{1}{\sqrt{2}\,v_0^2\, D^{3/2}\, t^{7/2}}\,
{\tilde P} \left( \frac{v_0 t- x}{v_0\, D\,  t^2}, \frac{y}{v_0\, \sqrt{2\,D}\, t^{3/2}}\right)\, ,
\label{typical_scaling}
\end{equation}
where the double Laplace transform of the scaling function ${\tilde P}(a_1,a_2)$ was computed
exactly~\cite{Basu2018}. Inverting explicitly this Laplace transform to extract $P(x,y,t)$
in the typical regime of $x$ and $y$ is difficult and has not been performed up to now.
In contrast, the marginal distributions
$p_x(x,t)= \int dy\, P(x,y,t)$ and $p_y(y,t)= \int dx\, P(x,y,t)$ were explicitly computed in the
typical fluctuation regime, leading to the expressions
\begin{eqnarray}
p_x(x,t)&= &\frac{1}{v_0\, D\, t^2}\, f_x\left(\frac{v_0\, t-x}{v_0\, D\, t^2}\right)\,, \label{x_marg.st}\\
p_y(y,t)&= & \frac{1}{v_0\, \sqrt{2D}, t^{3/2}}\,
f_y\left(\frac{y}{v_0\, \sqrt{2D}\, t^{3/2}}\right)\, ,
\label{y_marg.st}
\end{eqnarray}
The scaling function $f_x(z)$, supported over $z\in [0,\infty]$, is highly non-Gaussian
and has the asymptotic tails~\cite{Basu2018}
\begin{eqnarray}
\label{fxz.1}
f_x(z) \simeq \begin{cases}
&\frac{1}{2\,\sqrt{\pi\, z^3}}\, e^{-1/(8\,z)} \quad {\rm as}\quad z\to 0  \,,\\
\\
&\frac{1}{\sqrt{2\,z}}\, e^{-\pi^2\, z/8} \quad {\rm as}\quad z\to \infty \, .
\end{cases}
\end{eqnarray}
In contrast, the scaling function for the $y$ marginal is a pure Gaussian~\cite{Basu2018}
\begin{equation}
f_y(z)= \sqrt{\frac{3}{2\, \pi}}\, e^{-3 z^2/2} \, , \quad  z\in [-\infty,\infty]\, .
\label{fyz.1}
\end{equation}
However, the \emph{tails} of the marginal distributions $p_x(x,t)$, $p_y(y,t)$, and the joint distribution
$P(x,y,t)$ that characterize the atypical fluctuations of the positions $x$ and $y$, on a scale
$x\sim v_0\, t$ and $y\sim v_0\, t$ at short times $t\ll D^{-1}$, are yet to be understood.

In this paper, we derive these large deviation tails
by using
the optimal fluctuation method. Our main results can be summarized
as follows. For the marginal distributions we show that
\begin{eqnarray}
-\ln p_x(x,t) &\simeq & \frac{1}{2 D t}\, s_X\left(X= \frac{x}{v_0\, t}\right)\,,  \label{x.rescaledaction} \\
-\ln p_y(y,t) &\simeq & \frac{1}{2 D t}\, s_Y\left(Y= \frac{y}{v_0\, t}\right)\,,  \label{y.rescaledaction}
\end{eqnarray}
where the two rate functions $s_X(X)$ and $s_Y(Y)$ are computed analytically in Eqs.~\eqref{Xvsa} and~\eqref{svsa},  and Eqs.~\eqref{SY_alt.2} and~\eqref{S_action.1}, respectively.  We show that when $X\ll 1$ and
$Y\ll 1$, these large deviation tails match smoothly with the typical behaviors
in Eq.~\eqref{x_marg.st} and~\eqref{y_marg.st},  respectively. At the edges of their supports, $x=-v_0 t$ and $y=\pm v_0 t$, respectively, the marginal distributions vanish extremely rapidly and exhibit essential singularities, see Eqs.~(\ref{neartailx}) and (\ref{essentialy}). Equations~(\ref{x.rescaledaction}) and~(\ref{y.rescaledaction})  are asymptotically exact in the limit of $Dt\to 0$.

The joint distribution $P(x,y,t)$ exhibits the large deviation form
\begin{equation}
-\ln P(x,y,t) = \frac{1}{2 D t}\, s\left(X=  \frac{x}{v_0\,t}, Y= \frac{y}{v_0\, t}\right)\, . \label{rescaledaction}
\end{equation}
In a close vicinity of the point $(X=1,Y=0)$ where the joint distribution has its maximum, we show
that
the
rate function $s(X,Y)$ has an interesting self-similar structure:
\begin{equation}
s(X,Y) \simeq (1-X)\, F\left( \frac{1-X}{Y^2}\right)\, ,
\label{sxy_scaling}
\end{equation}
which follows from invariance of the governing equations in this limit under a stretching transformation.
We computed the scaling function $F(z)$ analytically, see Eq.~\eqref{summaryF}, and
plotted it in Fig.~\ref{fig:linearized}.  Similarly to the marginal distributions, this large deviation
tail matches smoothly
with the typical behavior of the
joint distribution. Close to the point $(X=1,Y=0)$ Eqs.~(\ref{rescaledaction}) and~(\ref{sxy_scaling}) become asymptotically exact in the limits of $Dt\to 0$, $1-X\to 0$ and $Y\to 0$ while keeping the ratio $z=(1-X)/Y^2$ constant.

The rest of the paper is organized as follows. In Section II we briefly introduce the optimal fluctuation method (OFM)
and formulate the OFM problem for the position statistics of the ABP. The complete marginal distributions are calculated in Section III.
In Section IV we study the joint distribution. Section V contains a brief
summary and discussion. Some technical details are relegated to the three appendices.

\section{The short-time position statistics and the OFM}

We consider the ABM model in two dimensions, defined in Eqs.~\eqref{xdoteq}-\eqref{phidot}, with the initial conditions
$x(0)=y(0)=0$ and $\phi(0)=0$. Let us first establish some general facts. Purely on dimensional ground, the
joint distribution $P(x,y,t)$, for arbitrary $t$, can be written as
\begin{equation}\label{Pexact}
P(x,y,t)= \frac{1}{v_0^2 t^2} \,\Phi \left(X,Y,Dt \right),
\end{equation}
where $X=x/(v_0 t)$ and $Y=y/(v_0 t)$ are the rescaled coordinates, and
\begin{equation}\label{normalization}
\int_{-\infty}^{\infty} \int_{-\infty}^{\infty} dX \, dY \Phi(X,Y,\,D\,t) =1.
\end{equation}
In their turn, the exact marginal $x$- and $y$-distributions scale as
\begin{equation}
p_x(x,t)= \int_{-\infty}^{\infty} P(x,y,t) dy = \frac{1}{v_0t} \, f\left(X,Dt\right),
\label{x_marg.1}
\end{equation}
and
\begin{equation}
p_y(y,t) = \int_{-\infty}^{\infty} P(x,y,t) dx = \frac{1}{v_0t}\, g\left(Y,Dt\right),
\label{y_marg.1}
\end{equation}
where
\begin{eqnarray}
  f(X,Dt) &=& \int_{-\infty}^{\infty} \Phi(X,Y,Dt)\, dY, \label{margx}\\
  g(Y,Dt)&=& \int_{-\infty}^{-\infty} \Phi(X,Y,Dt)\, dX. \label{margy}
\end{eqnarray}
As argued in the introduction, the exact distribution $P(x,y,t)$ vanishes
outside the circle of unit radius $X^2+Y^2=1$.
Consequently, the marginal $x$ and $y$ distributions live on the intervals
$|X|\le 1$ and $|Y|\le 1$, respectively.

Here we will study the short-time behavior, $D t\ll 1$, of $p_x(x,t)$, $p_y(y,t)$ and $P(x,y,t)$.
Since we are interested
in the large deviation regime $X\sim O(1)$ and $Y\sim O(1)$, we can employ the OFM.

\vspace{0.5 cm}

\subsection{Optimal fluctuation method}
\label{OFM}

At short times we can calculate $P(x,y,t)$ up to a preexponential factor by applying the
OFM which, for the Brownian motion, becomes geometrical optics ~\cite{GF2003,Meerson2019,SmithMeerson2019,MeersonSmith2019,Agranovetal2020,MM2020}. The OFM
boils down to a minimization of the action $S$ of the Brownian motion of the angle $\phi(t')$,
\begin{equation}\label{action}
-\ln P(x,y,t) \simeq S = \frac{1}{4D}\int_0^t  {\dot \phi}^2 (t')\, dt',
\end{equation}
over all paths $\phi(t')$ subject to the integral constraints
\begin{equation}\label{condxy}
v_0 \int_0^t \cos\phi(t') dt' =x\,, \quad
v_0 \int_0^t \sin \phi(t') dt' =y.
\end{equation}
The boundary conditions for $\phi(t')$, for $0\le t'\le t$, are
\begin{equation}\label{BCs}
\phi(t'=0)=0 \quad \text{and} \quad \dot{\phi}(t'=t) = 0,
\end{equation}
where the second condition is the free-end boundary condition of the calculus of variations \cite{Elsgolts}.
Going over to the rescaled variables $X=x/(v_0t)$, $Y=y/(v_0t)$ and ${\tau}=t'/t$,  we immediately arrive at the
OFM scaling behavior, announced in Eq.~\eqref{rescaledaction}
[compare it with the exact scaling in \eqref{Pexact}].
The rescaled OFM action
\begin{equation}\label{s}
s(X,Y)= \frac{1}{2} \int_0^1 {\dot {\phi}}^2(\tau)\, d\tau
\end{equation}
plays the role of a rate function. The rescaled integral conditions~(\ref{condxy}) become
\begin{equation}\label{condX}
\int_0^1 \cos \phi(\tau) d\tau =X
\end{equation}
and
\begin{equation}\label{condY}
\int_0^1 \sin \phi(\tau) d\tau =Y \, .
\end{equation}
The two boundary conditions in \eqref{BCs}
become
\begin{equation}
\phi(\tau=0)=0 \quad \text{and} \quad \dot{\phi}(\tau=1) = 0\,.
\label{BCs_tau}
\end{equation}

\section{Marginal distributions}
We first study the short-time behavior of the  marginal distributions
$p_x(x,t)$ and $p_y(y,t)$.

\subsection{Marginal distribution $p_x(x,t)$}
\label{xdist}

Within the OFM framework, the marginal $x$-distribution has the form \eqref{x.rescaledaction},
where the rescaled action $s_X(X)$ should be evaluated
by accounting only for the constraint (\ref{condX}).
Introducing a Lagrange multiplier $\lambda$ to enforce this constraint, we can write the effective action as
\begin{equation}
s_X(X)= \frac{1}{2} \int_0^1 {\dot{\phi}}^2(\tau)\, d\tau - \lambda \left(X-\int_0^1 \cos \phi(\tau)\, d\tau\right)\, .
\label{eff_action_X.1}
\end{equation}
This action corresponds to the Lagrangian of a Newtonian particle in an effective potential
$V(\phi)= -\lambda\, \cos \phi$. The Euler-Lagrange equation reads
\begin{equation}
\ddot{\phi} + \frac{\partial V}{\partial \phi}=\ddot{\phi} +\lambda \sin \phi =0\, .
\label{EL_X.1}
\end{equation}
This is the pendulum equation. Its energy integral is
\begin{equation}\label{energyintegral}
\frac{\dot{\phi}^2}{2} - \lambda \cos \phi= E =\text{const}\, .
\end{equation}
In view of the boundary conditions~(\ref{BCs_tau}), the pendulum must start at $\tau=0$ from $\phi=0$
and stop at $\tau=1$. The least-action solution (the optimal path) is
such that $\phi(\tau)$ is a monotonic function on the interval $0<\tau<1$, and we can choose it to be
monotone increasing~\cite{footnote1}.
Therefore, \eqref{energyintegral} yields
\begin{equation}\label{tvsphi}
\int_0^{\phi(\tau)} \frac{dz}{\sqrt{2\lambda(a+\cos z)}} = \tau .
\end{equation}
Here we have already accounted for the condition $\phi(0)=0$ and introduced the parameter $a=E/\lambda$, so
that $|a|<1$. The stopping point of the pendulum at $\tau=1$ is at $\phi(\tau=1)\equiv \phi_*=\arccos(-a)$, and
$0<\phi_*<\pi$. Setting $\tau=1$ in \eqref{tvsphi}, we obtain a relation between $\lambda$ and $a$:
\begin{equation}\label{lambdavsa}
\sqrt{2\lambda} = \int_0^{\phi_*} \frac{dz}{\sqrt{a+\cos z}} =\frac{2}{\sqrt{a+1}}\, F\left(\frac{\phi_*}{2},\frac{2}{a+1}\right),
\end{equation}
where $F(\phi,m)=\int_0^{\phi} (1-m \sin^2 z)^{-1/2} dz$ is the elliptic integral of the first kind.
The constraint (\ref{condX}) yields a relation between $a$ and $X$:
\begin{eqnarray}
  \!\!\!\!\!\!\!\!\!\!\!\!&&X=\int_0^1 \cos \phi(\tau)\, d\tau =
\int_0^{\phi_*} \!\!\!\cos \phi\, \frac{d\tau}{d\phi} \,d\phi \nonumber\\
 \!\!\!\!&&= \int_0^{\phi_*}\!\!\frac{\cos z\,dz}{\sqrt{2\lambda\,(a+\cos z)}} \!=\! \frac{(a+1) E\left(\frac{\phi_*}{2},\frac{2}{a+1}\right)}{F\left(\frac{\phi_*}{2},
  \frac{2}{a+1}\right)}-a,
  \label{Xvsa}
\end{eqnarray}
where  $E(\phi,m)=\int_0^{\phi} (1-m \sin^2 z)^{1/2}\, dz$ is
the elliptic integral of the second kind, and we have used \eqref{lambdavsa}.
Now we can express the action $s$ in \eqref{eff_action_X.1} via $a$.
Using the energy integral~(\ref{energyintegral}), we obtain
\begin{eqnarray}
  s_X(a) &=& {\lambda} \int_0^1  \left(a+\cos \phi\right)\, d\tau =
{\lambda}\, (a+X)\nonumber\\
  &=&  2\, F\left(\frac{\phi_*}{2} ,\frac{2}{a+1}\right)\, E\left(\frac{\phi_*}{2},\frac{2}{a+1}\right),
  \label{svsa}
\end{eqnarray}
where we have used Eqs.~(\ref{condX}), (\ref{lambdavsa}) and (\ref{Xvsa}).

\begin{figure}[ht]
\includegraphics[width=0.35\textwidth,clip=]{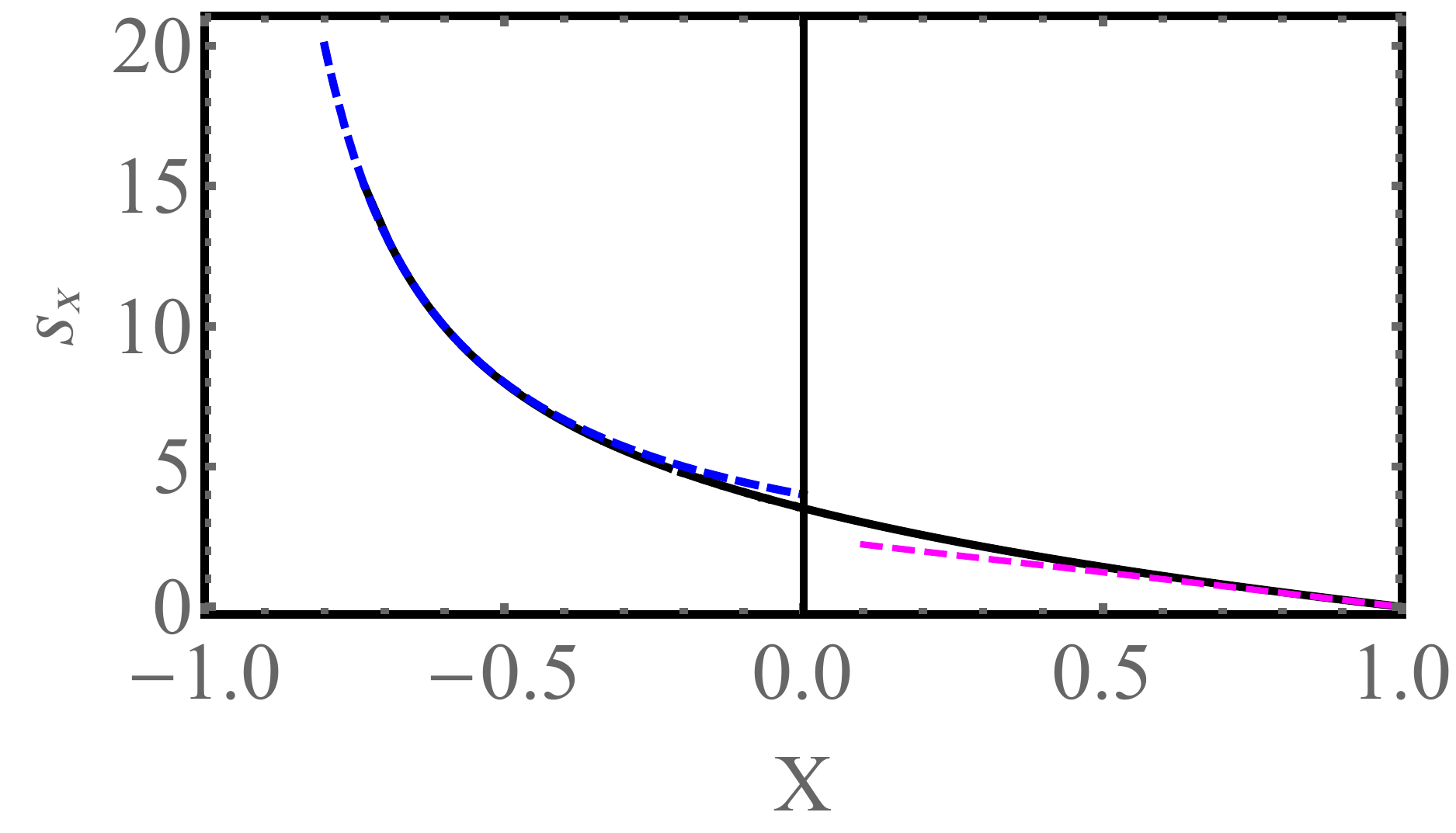}
\caption{
Solid line: the rescaled action $s_X(X)$, given by  Eqs.~(\ref{Xvsa}) and (\ref{svsa}). The dashed lines are the asympto- tics (\ref{asympX}).}
\label{actionXfig}
\end{figure}

Equations~(\ref{Xvsa}) and (\ref{svsa}) determine the action $s=s_X(X)$ in the parametric form
$X=X(a), \;s=s_X(a)$. A plot of $s=s_X(X)$ is shown in Fig. \ref{actionXfig}.
Also shown are two asymptotes of $s_X(X)$, for $1-X\ll 1$ and for $1+X\ll 1$:
\begin{eqnarray}
s_X(X)\simeq \begin{cases}
& \frac{\pi^2}{4}\left(1-X\right),
\quad \quad 1-X\ll 1, \\
\\
& \frac{4}{1+X},\quad \quad \quad \quad\quad 1+X\ll 1 .
\end{cases}
\label{asympX}
\end{eqnarray}
The upper line in \eqref{asympX} correspond to $1+a\ll 1$, the lower line to $1-a\ll 1$.

\vspace{0.5cm}
Equation~(\ref{x.rescaledaction}) with the OFM rate function, defined parametrically
by Eqs.~(\ref{Xvsa}) and (\ref{svsa}), gives a complete description of large deviations of $x$.
The asymptotic $1-X\ll 1$ from \eqref{asympX} describes an exponential tail of $p_x(x,t)$,
\begin{equation}\label{neartailx}
-\ln p_x(x,t) \simeq \frac{\pi^2 (v_0 t -x)}{8 D v_0 t^2},
\quad D v_0 t^2\ll v_0t-x\ll v_0t .
\end{equation}
The left strong inequality guarantees that the action~(\ref{action}) is large, $S\gg 1$, as required for the
applicability of the OFM. The ``near tail" (\ref{neartailx}) agrees with the tail of
of the typical behavior of $p_x(x,t)$  in the second line of Eq.~\eqref{fxz.1}.

The asymptotic $1+X\ll 1$ in \eqref{asympX} describes an essential singularity of the
marginal distribution $p_x(x,t)$ at the leftmost point of its support,
$x=-v_0t$, where the distribution vanishes, together with all its $x$-derivatives,  extremely fast:
\begin{equation}\label{essentialx}
-\ln p_x(x\to -v_0t,t) \simeq \frac{2 v_0}{D (v_0t+x)} .
\end{equation}

\subsection{Marginal distribution $p_y(y,t)$}
\label{ydist}

\begin{figure}[ht]
\includegraphics[width=0.35\textwidth,clip=]{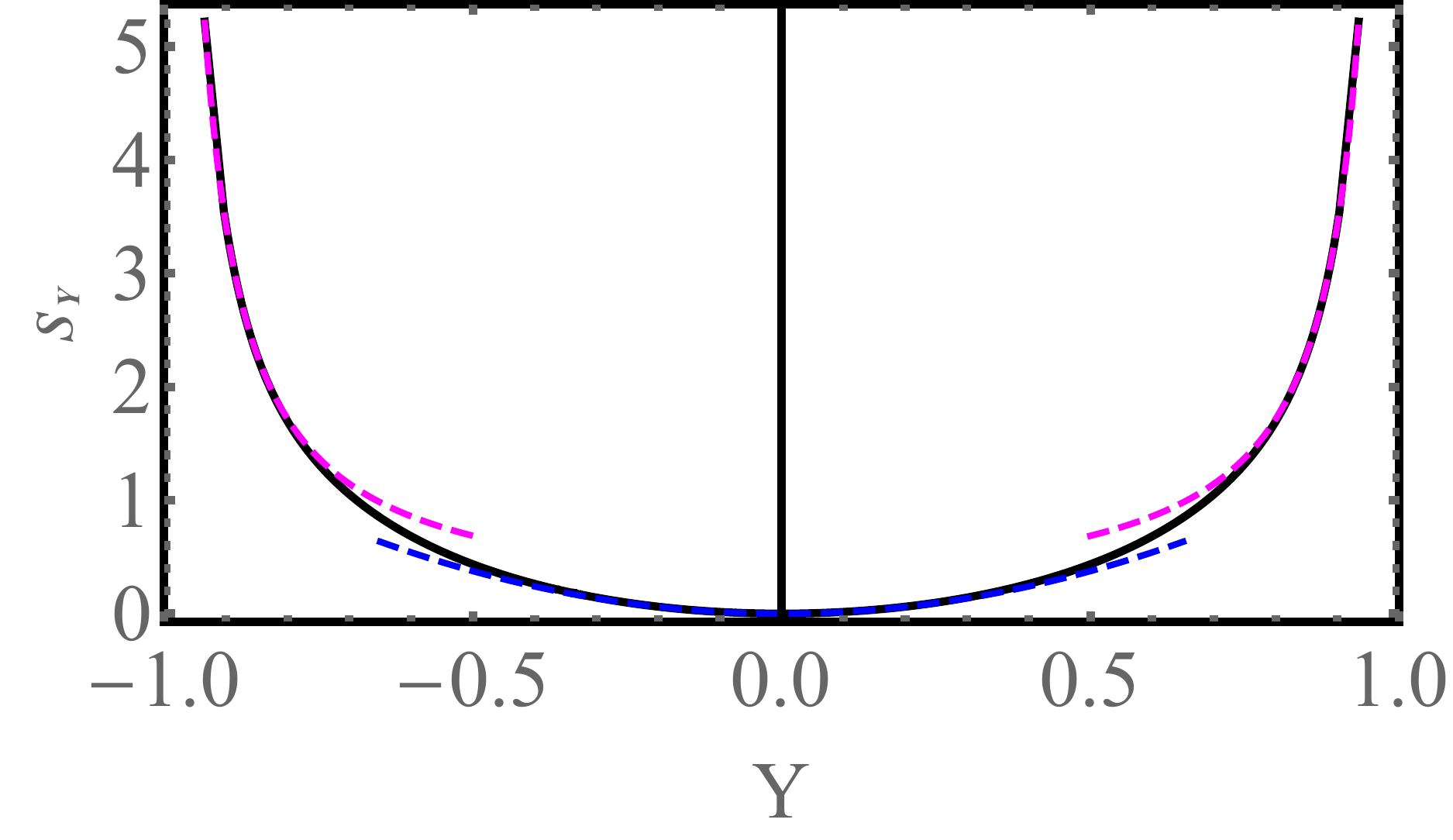}
\caption{
Solid line: the rescaled action $s_Y(Y)$, given by  Eqs.~(\ref{Xvsa}) and (\ref{svsa}).
The dashed lines are asymptotics (\ref{asympY}).}
\label{actionYfig}
\end{figure}

In this case the OFM ansatz takes the form Of Eq.~\eqref{y.rescaledaction},
where the rescaled action $s_Y(Y)$ should be evaluated
by accounting only for the constraint (\ref{condY}).
Introducing a proper Lagrange multiplier $\lambda$,
we can write the effective rescaled action as
\begin{equation}
s_Y(Y)= \frac{1}{2} \int_0^1 {\dot{\phi}}^2(\tau)\, d\tau + \lambda \left(Y-\int_0^1 \sin \phi(\tau)\, d\tau\right)\, .
\label{eff_action_Y.1}
\end{equation}
This is the Lagrangian of a Newtonian particle in an effective potential
$V(\phi)= \lambda\, \sin \phi$. The
Euler-Lagrange equation reads
\begin{equation}
\ddot{\phi} + \frac{\partial V}{\partial \phi} = \ddot{\phi} +\lambda \cos \phi =0,
\label{EL_Y.1}
\end{equation}
with the energy integral
\begin{equation}\label{energyintegralY}
\frac{\dot{\phi}^2}{2} + \lambda\, \sin \phi= E \, .
\end{equation}
As the $y$-distribution is symmetric with respect to $y=0$, it suffices to consider $y>0$. Here $\phi(\tau)$
increases monotonically from $\phi=0$ until the pendulum stops at $\tau=1$. The calculations are similar to those
for the $p_x(x,t)$. We define, as before, $a=E/{\lambda}$ (note that $|a|<1$) and integrate Eq.~\eqref{energyintegralY},
using the boundary conditions $\phi(\tau=0)=0$ and ${\dot \phi}(\tau=1)=0$. This gives
\begin{equation}
\sqrt{2\lambda}= \int_0^{\phi(1)} \frac{d\phi}{\sqrt{a-\sin \phi}}\, ,
\label{2l_Y}
\end{equation}
where $\sin\phi(1)=a$. Defining $\phi_*= \arccos(-a)$, we obtain
$\phi(1)=\phi_*-\pi/2$. Making a change of variable $\phi=z-\pi/2$,
one can express the integral  in Eq.~\eqref{2l_Y} in terms of the elliptic function $F(\phi,m)$, defined earlier. We obtain
\begin{equation}
\sqrt{2\lambda}= \frac{2}{\sqrt{a+1}}\,\left[F\left(\frac{\phi_*}{2},
\frac{2}{a+1}\right)- F\left(\frac{\pi}{4},\frac{2}{a+1}\right)\right]\, .
\label{2l_Y.1}
\end{equation}
Using Eq.~\eqref{energyintegralY}, we express the constraint~\eqref{condY} as
\begin{eqnarray}
Y & = & \int_0^1 \sin\phi(\tau)\, d\tau= \int_0^{\phi(1)} \sin \phi(\tau)\, \frac{d\tau}{d\phi}\, d\phi \nonumber \\
&=& \frac{1}{\sqrt{2\lambda}}\, \int_0^{\phi(1)}  \frac{\sin \phi\, d\phi}{\sqrt{a-\sin \phi}} \nonumber \\
&=& \frac{1}{\sqrt{2\lambda}}\left[a- \int_0^{\phi(1)} \sqrt{a-\sin \phi}\, d\phi \right]\, ,
\label{Y1.1}
\end{eqnarray}
where we used the identity $\sin \phi=a-(a-\sin \phi)$ in arriving at the third line from the second line.
Making the same change of variable $\phi= z-\pi/2$, we can express the integral in terms of the
elliptic function $E(\phi,m)$. Finally, eliminating $\sqrt{2\lambda}$ from Eqs. (\ref{Y1.1})
and (\ref{2l_Y.1}), we obtain
\begin{equation}
Y= a- (a+1)\, \frac{E\left(\frac{\phi_*}{2},
\frac{2}{a+1}\right)- E\left(\frac{\pi}{4},\frac{2}{a+1}\right)}{F\left(\frac{\phi_*}{2},
\frac{2}{a+1}\right)- F\left(\frac{\pi}{4},\frac{2}{a+1}\right)}\, .
\label{SY_alt.2}
\end{equation}
The action $s$ in \eqref{eff_action_Y.1} can also be evaluated:
\begin{eqnarray}
s_Y(a) & = & \lambda (a-Y) \nonumber \\
& =& 2\, \left[F\left(\frac{\phi_*}{2},
\frac{2}{a+1}\right)- F\left(\frac{\pi}{4},\frac{2}{a+1}\right)\right] \nonumber \\
&\times&\left[E\left(\frac{\phi_*}{2},
\frac{2}{a+1}\right)- E\left(\frac{\pi}{4},\frac{2}{a+1}\right)\right]\, ,
\label{S_action.1}
\end{eqnarray}
where we used Eq.~\eqref{2l_Y.1} for $\lambda$ and Eq.~\eqref{SY_alt.2} for $Y$.
Again, we obtained $s_Y(Y)$ in a parametric form,
determined by Eqs. (\ref{SY_alt.2}) and (\ref{S_action.1}). A plot of $s_Y(Y)$ vs. $Y$, is shown in Fig.~\ref{actionYfig}.
One can also derive the following asymptotic behaviors
of $s_Y(Y)$ for $|Y|\ll 1$ and $1-Y\ll 1$:
\begin{eqnarray}
s_Y(Y)\simeq \begin{cases}
& (3/2) Y^2, \quad \quad \quad \quad\quad\quad |Y| \ll 1, \\
\\
& \frac{2(3-2\sqrt{2})}{1-|Y|},\quad \quad \quad \quad\quad 1-|Y|\ll 1 .
\end{cases}
\label{asympY}
\end{eqnarray}
These two asymptotics are also plotted in Fig. \ref{actionYfig}.

Using the asymptotic behavior \eqref{asympY} for $|Y|\ll 1$ in Eq.~\eqref{y.rescaledaction}, and normalizing to unity,
we obtain a Gaussian distribution which describes \emph{typical} fluctuations of $y$, that is small distances, $|y|\ll v_0 t$:
\begin{equation}\label{Gaussy}
p_y(y,t) \simeq \frac{1}{\sqrt{2\pi \sigma_y^2}}\,e^{-\frac{y^2}{2\sigma_y^2}}, \quad \text{where}\quad \sigma_y^2 = \frac{2}{3}v_0^2 Dt^3,
\end{equation}
This distribution  is in perfect
agreement with the typical behavior in \eqref{fyz.1}.
The complete $y$-distribution, however,  is strongly non-Gaussian, as one can see from Eq.~\eqref{S_action.1}. In particular, it vanishes extremely rapidly at the leftmost and rightmost points of its support, $y=\pm v_0t$:
\begin{equation}\label{essentialy}
-\ln p_y(|y|\to v_0t,t) \simeq \frac{(3-2\sqrt{2}) v_0}{D (v_0t-|y|)} ,
\end{equation}
and has an essential singularity there, which is similar to that of the $x$-distribution at $x=v_0t$, see
Eq.~\eqref{essentialx}.

\section{Joint distribution $P(x,y,t)$ }
\label{joint_P}

Now we consider the joint distribution $P(x,y,t)$ in the $(X,Y)$ plane
where $X=x/(v_0t)$ and $Y=x/(v_0t)$ are the rescaled co-ordinates.
As we showed in the Introduction, the joint distribution is supported
on the circle $X^2+Y^2=1$ of unit radius. For the initial condition
$\phi(0)=0$, the distribution is peaked at $(X=1,Y=0)$ and vanishes
at all other points on the circumference of the circle.
At short times $P(x,y,t)$ obeys the OFM scaling behavior (\ref{rescaledaction}),
and the rescaled action $s(X,Y)$ can be written as
\begin{eqnarray}
s(X,Y) &= &\frac{1}{2} \int_0^1 {\dot{\phi}}^2(\tau)\, d\tau-
\lambda_1\, \left(X-\int_0^1 \cos \phi(\tau)\, d\tau\right) \nonumber \\
& + &\lambda_2\, \left(Y-\int_0^1 \sin \phi(\tau)\, d\tau\right)\, .
\label{eff_action_XY.1}
\end{eqnarray}
The two Lagrange multipliers $\lambda_1$ and $\lambda_2$ enforce the two constraints
in Eqs. (\ref{condX}) and (\ref{condY}), respectively. In principle, one can
determine the complete joint distribution in this formalism. This would require solving the effective mechanical problem,
defined by the action~(\ref{eff_action_XY.1}), exactly.  The exact solution can be obtained in elliptic functions.
In this paper, however, we
restrict ourselves for simplicity to a close vicinity of the point $(X=1,Y=0)$ where
the probability density is the largest. Near this point we can use a small-angle approximation, which considerably
simplifies the OFM analysis.

Indeed,  when $X$ is very close to $1$, and $Y$ is small,
it is clear from Eqs. (\ref{condX}) and (\ref{condY}),
that $\phi(\tau)$ stays close to $0$ for all $0\le \tau\le 1$. This justifies
the small-angle approximation,
$\cos \phi\simeq 1-\phi^2/2$ and $\sin \phi \simeq \phi$.
In this approximation
the rescaled OFM action in \eqref{eff_action_XY.1} becomes
\begin{eqnarray}
s(X,Y) &\simeq & \frac{1}{2} \int_0^1 {\dot{\phi}}^2(\tau)\, d\tau-
\lambda_1  \left(X-1+ \frac{1}{2}\int_0^1 \phi^2(\tau)\, d\tau\right) \nonumber \\
& + & \lambda_2  \left(Y-\int_0^1 \phi(\tau)\, d\tau\right)
\label{eff_action_XY.2}
\end{eqnarray}
subject to the constraints
\begin{eqnarray}
X &= & 1- \frac{1}{2} \int_0^1 \phi^2(\tau)\, d\tau \,,\label{X_cons} \\
Y &=& \int_0^1 \phi(\tau)\, d\tau \,.\label{Y_cons}
\end{eqnarray}
The approximate effective action $s(X,Y)$ in Eq.~\eqref{eff_action_XY.2} corresponds to the
Lagrangian of a Newtonian particle in a parabolic potential
\begin{equation}
V(\phi)= \frac{\lambda_1}{2}\, \phi^2 + \lambda_2\, \phi\, .
\label{harmonic.1}
\end{equation}
The Euler-Lagrange equation for the optimal path $\phi(t)$ reads
\begin{equation}
\ddot{\phi}= -\frac{\partial V(\phi)}{\partial \phi}= -\lambda_1\,
\phi- \lambda_2= -\lambda_1 (\phi + B)\, ,
\label{EL.1}
\end{equation}
where $B= \lambda_2/\lambda_1$. The solution is elementary, but with a twist:  It turns out that there are two branches of the solution, depending on the sign of $\lambda_1$. Let us consider the two cases separately.

\begin{figure*}[ht]
    \begin{subfigure}[b]{0.31\textwidth}
        \includegraphics[width=\textwidth]{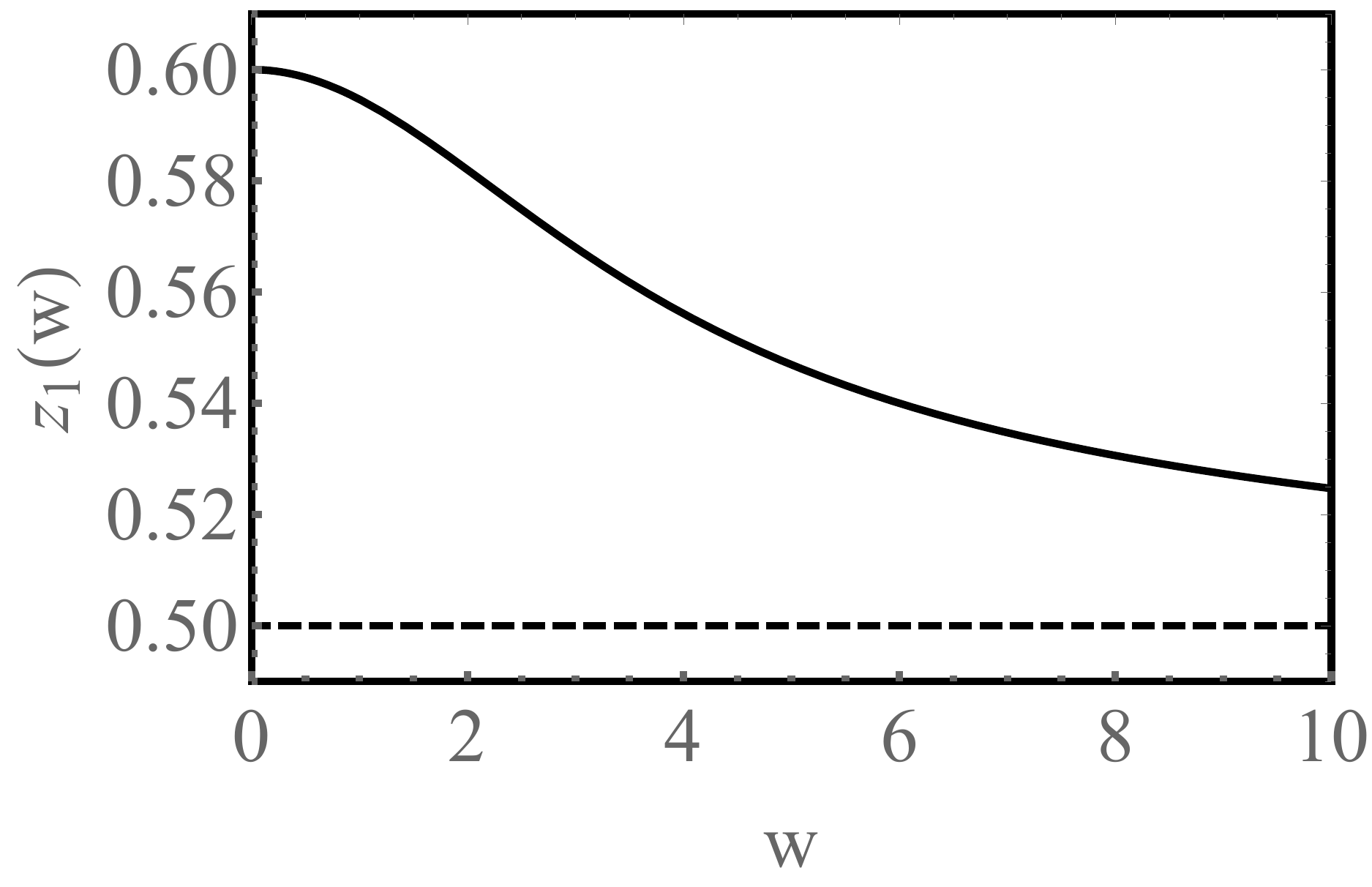}
        \caption{}
        \label{fig:zw1}
    \end{subfigure}
\quad
    \begin{subfigure}[b]{0.31\textwidth}
        \includegraphics[width=\textwidth]{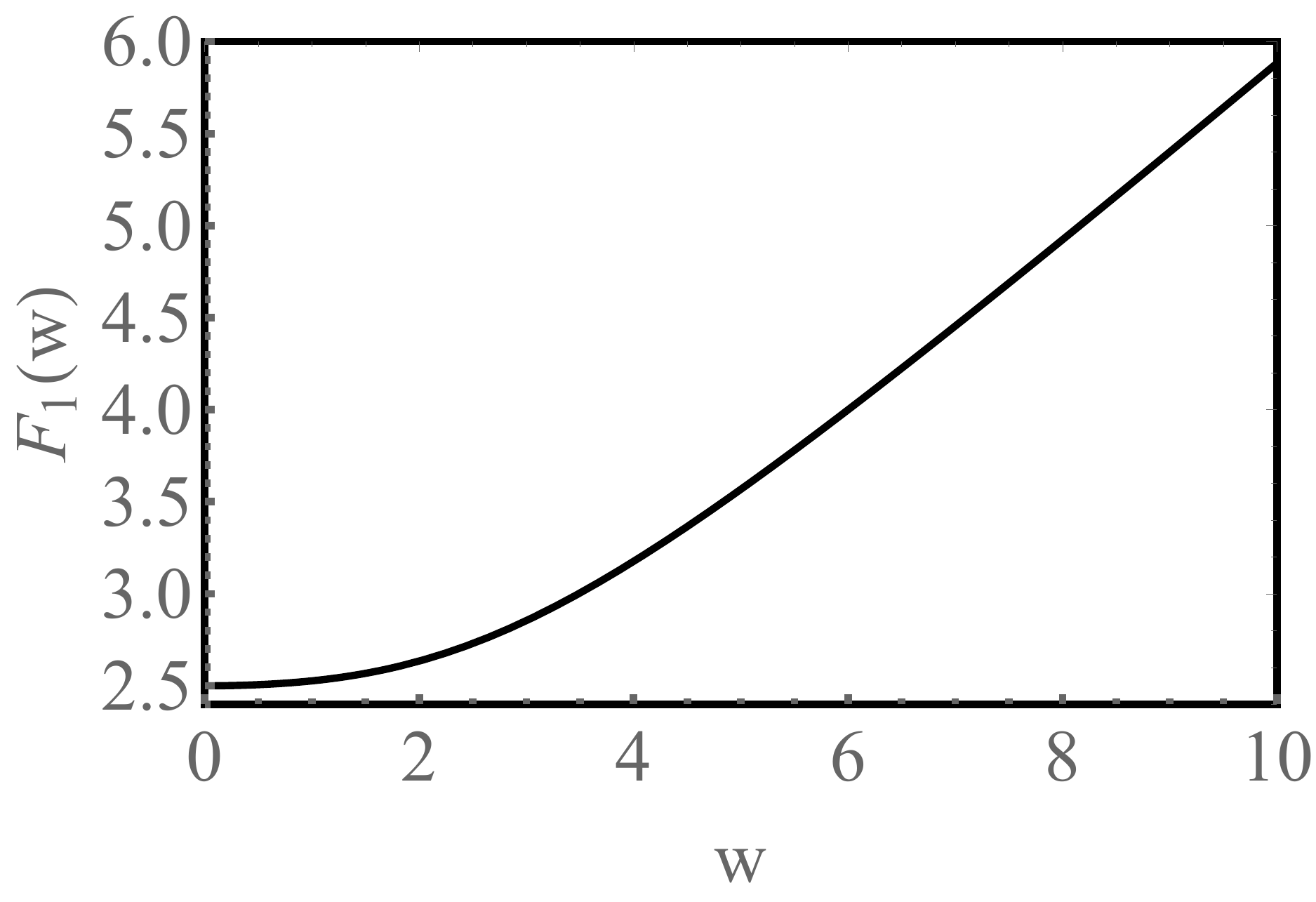}
        \caption{}
        \label{fig:Fw1}
    \end{subfigure}
\quad
    \begin{subfigure}[b]{0.31\textwidth}
        \includegraphics[width=\textwidth]{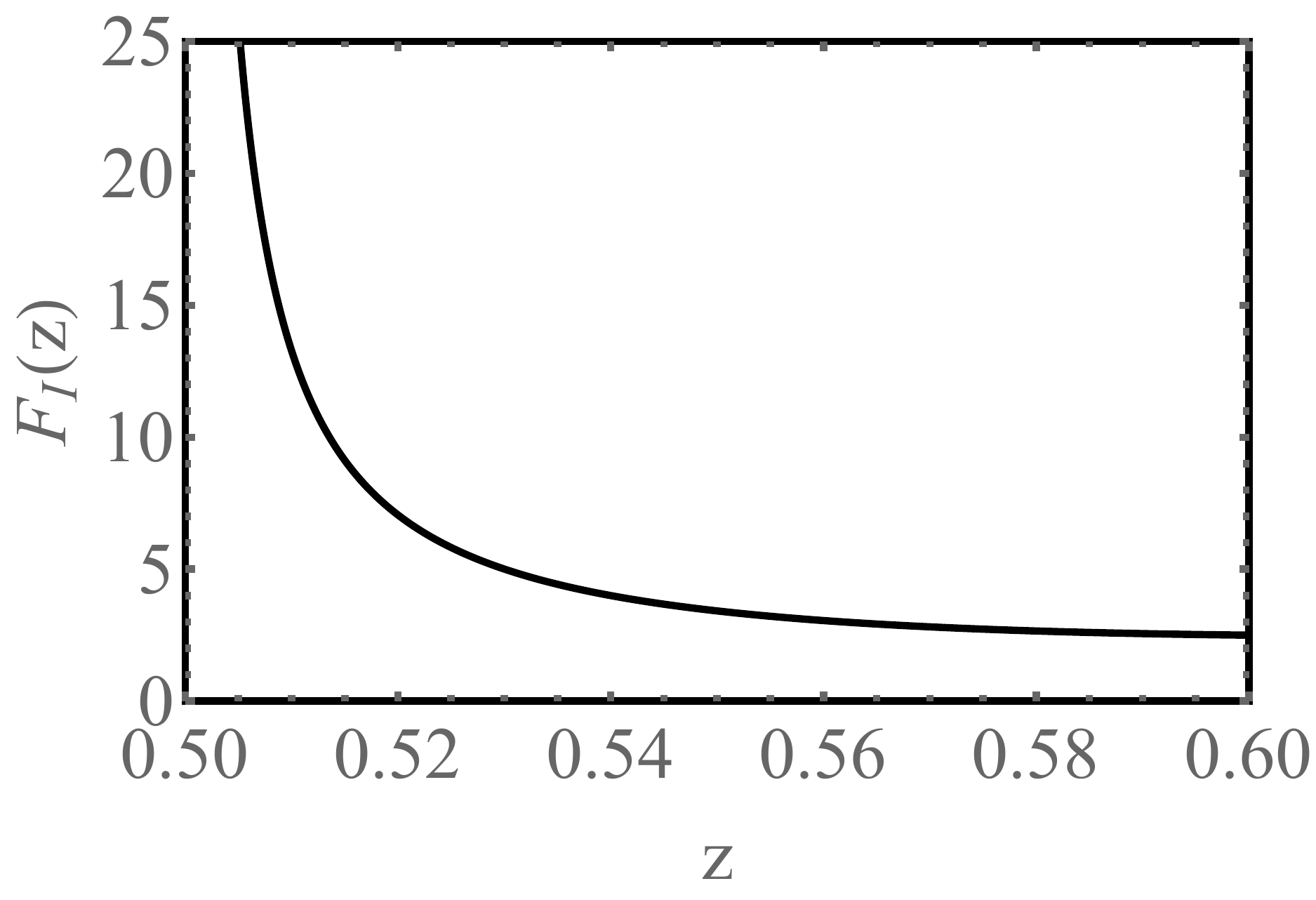}
        \caption{}
        \label{fig:Fz1}
    \end{subfigure}
\caption{(a) The function $z_{1}(w)$ vs. $w$ from \eqref{zw_2.1}.
As $w$ increases, $z_{1}(w)$ decreases monotonically from $z_{1}(0)=3/5$
and approaches $z_{1}=1/2$ as $w\to \infty$.
(b) The function $F_{1}(w)$ vs. $w$ from \eqref{Fz_2.1}.
It increases monotonically with $w$ starting from $F_{1}(0)=5/2$.
Asymptotically, $F_{1}(w)\to w/2$ as $w\to \infty$. (c) The scaling function $F_{\rm I}(z)$ vs. $z$,
defined for $1/2\le z\le 3/5$ and determined parametrically by
Eqs. ~\eqref{Fz.1} and \eqref{zw.1}. When $z\to 3/5$, $F_{\rm I}(z)\to 5/2$, while
for $z\to 1/2$, $F_{\rm I}(z)$ diverges as
$[8\,(z-1/2)]^{-1}$.}
\label{fig:branchI}
\end{figure*}

\subsubsection{$\lambda_1\le 0$: Branch I}
For $\lambda_1\le 0$ the particle
moves in an \emph{inverted} parabolic potential. Let us set
$\lambda_1= - w^2$ with $w\ge 0$. Solving Eq.~\eqref{EL.1} with the boundary conditions
$\phi(\tau=0)=0$ and ${\dot \phi}(\tau=1)=0$, we obtain
\begin{equation}
\phi(\tau)= B \,\left[\frac{\cosh\left(w (\tau-1)\right)}{\cosh w} - 1\right]\,, \quad 0\le \tau\le 1\,.
\label{solI.1}
\end{equation}
This solution is monotonic on the whole time interval $\tau\in [0,1]$:
the Newtonian particle moves in one direction. Indeed, the time derivative
\begin{equation}
{\dot \phi} (\tau)= B\, w\, \frac{\sinh\left[w (\tau-1)\right]}{\cosh w}
\label{dotphiI.1}
\end{equation}
does not change sign on the interval $0\le \tau<1$ and vanishes only at the end of the interval $\tau=1$.

The two constants $w$ and $B$ in \eqref{solI.1} are just reparametrizations of
$\lambda_1$ and $\lambda_2$, and they have to be determined from the two constraints
\eqref{X_cons} and \eqref{Y_cons}. Plugging the solution (\ref{solI.1}) into these constraints gives
\begin{eqnarray}
1-X &= & \frac{B^2}{4}\, \left[ 3 \left(1- \frac{\tanh w}{w}\right)- \tanh^2 w
\right]\,, \label{X_cons_I.1} \\
Y &= & B\, \left(\frac{\tanh w}{w}-1\right)\, . \label{Y_cons_I.1}
\end{eqnarray}
Eliminating $B$ from Eqs.~(\ref{X_cons_I.1}) and (\ref{Y_cons_I.1}), we determine $w$ in terms of $X$ and $Y$:
\begin{equation}
\frac{1-X}{Y^2}=\frac{ 3 \left(1- \tanh w/w\right)- \tanh^2 w}{4\, \left(\tanh w/w-1\right)^2}\, .
\label{ratioI.1}
\end{equation}
Furthermore, by substituting the solution \eqref{solI.1}) in the action \eqref{eff_action_XY.2}
and eliminating $B^2$ using \eqref{X_cons_I.1}, we obtain
\begin{equation}
s(X,Y)= (1-X)\, \frac{w^2\, \left(\tanh^2 w+ \tanh w/w-1\right)}{3 \left(1- \tanh w/w\right)- \tanh^2 w} \, .
\label{action_I.1}
\end{equation}
Therefore, the action has a self-similar structure:
\begin{equation}
s(X,Y)= (1-X)\,F_{\rm I}\left(\frac{1-X}{Y^2}\right)\,,
\label{action_I.scaling}
\end{equation}
where the scaling function $F_{\rm I}(z)$ is parametrically determined by the equations
\begin{eqnarray}
z=z_1(w) &= & \frac{3\,\left(1- \tanh w/w\right)-
\tanh^2 w}{4\,(1-\tanh w/w)^2}\,, \label{zw_2.1}\\
F_{\rm I}=F_{1}(w) &=& \frac{ w^2\, \left(\tanh^2 w+ \tanh w/w-1 \right)}{3
\,\left(1- \tanh w/w\right)-\tanh^2 w}\,. \label{Fz_2.1}
\end{eqnarray}
A plot of the function $z_{1}(w)$ vs. $w$ is shown in Fig.~\ref{fig:zw1}.  This function has the following asymptotic behaviors:
\begin{eqnarray}
z_1(w)\simeq \begin{cases}
& \frac{3}{5}- \frac{w^2}{175}+ \ldots
\quad \quad w\to 0\,, \\
\\
& \frac{1}{2} + \frac{1}{4w} + \ldots  \quad\quad w\to \infty\,.
\end{cases}
\label{asymp_z1w}
\end{eqnarray}
A plot of $F_{1}(w)$ vs. $w$ is shown in Fig.~\ref{fig:Fw2}, and the asymptotic behaviors are
\begin{eqnarray}
F_1(w)\simeq \begin{cases}
& \frac{5}{2}+\frac{w^2}{42}+ \ldots
\quad \quad w\to 0\,, \\
\\
& \frac{w}{2} + \frac{3}{4} + \ldots  \quad\quad w\to \infty\,.
\end{cases}
\label{asymp_F1w}
\end{eqnarray}
The scaling function $F_{\rm I}(z)$, obtained
parametrically from $F_1(w)$ and $z_1(w)$ in Eqs. (\ref{Fz_2.1}) and (\ref{zw_2.1}) in
the range $1/2\le z\le 3/5$, is plotted vs. $z$ in Fig.~\ref{fig:Fz1}. It has
the limiting behaviors
\begin{eqnarray}
F_{\rm I}(z)\simeq \begin{cases}
& \frac{1}{8\, (z-1/2)} +\frac{3}{4}+\ldots\,,
\quad \quad z\to 1/2^{+}\,, \\
\\
& \frac{5}{2} + \frac{175}{42}\, (\frac{3}{5}-z)+\ldots\,,  \quad\quad z\to \frac{3}{5}\,.
\end{cases}
\label{limits_FIz}
\end{eqnarray}
We will refer to this range of $z \in [1/2,3/5]$ as branch I. Here the optimal path $\phi(\tau)$ is
monotonic on the whole time interval $\tau\in [0,1]$.

\subsubsection{$\lambda_1\ge 0$: Branch II}
In this case we can set
$\lambda_1=\omega^2$ (with $\omega\ge 0$) and solve \eqref{EL.1}
with the boundary conditions
$\phi(\tau=0)=0$ and $\dot \phi(\tau=1)=0$. The solution,
\begin{equation}
\phi(\tau) = B \,\left[\frac{\cos\left(w (\tau-1)\right)}{\cos w} - 1\right]\,,\quad 0\le \tau\le 1\,,
\label{phi_sol.1}
\end{equation}
can be obtained by replacing $w=i\, \omega$ in
Eq.~\eqref{solI.1}. In contrast to the branch I, where the optimal path was always
monotonic, here one can have both monotonic and non-monotonic
paths depending on the range of $\omega$. To see this,  consider the
time derivative
\begin{equation}
\dot \phi(\tau)= B\, \omega\,  \frac{\sin [\omega\, (1-\tau)]}{\cos \omega}\, .
\label{phidotII.1}
\end{equation}
As $\tau$ increases from $0$, the first zero of $\dot \phi$ occurs when $\omega\, (1-\tau)=\pi$, \textit{i.e.}
at $\tau=\tau_c= 1-\pi/\omega$. If $\tau_c>1$, \textit{i.e.} $0<\omega <\pi$,  $\dot \phi$ does not vanish
in the interval $0\le \tau\le 1$, and the optimal path is monotonic. On the contrary, if $\omega>\pi$, we have
$0 \le \tau_c\le 1$ and the path is nonmonotonic: our Newtonian particle changes its direction of motion
at $\tau=\tau_c\le 1$ before reaching $\tau=1$.

The rest of calculations are very similar to the preceding case: one only needs to replace
$w$ by $i\, \omega$ everywhere. In particular, Eq.~\eqref{ratioI.1} is now replaced by
\begin{equation}
\frac{1-X}{Y^2}= \frac{3\,\left(1- \tan \omega/\omega\right)+
\tan^2 \omega}{4\,(1-\tan \omega/\omega)^2}\, .
\label{omega.1}
\end{equation}
As a result, the action $s(X,Y)$ has the same self-similar form as in \eqref{action_I.1},
\begin{equation}
s(X,Y)= (1-X)\,F_{\rm II}\left(\frac{1-X}{Y^2}\right)\,,
\label{action.3}
\end{equation}
but the scaling function $F_{\rm II}(z)$ is now parametrically determined by the equations
\begin{eqnarray}
z=z_{2}(\omega) &= & \frac{3\,\left(1- \tan \omega/\omega\right)+
\tan^2 \omega}{4\,(1-\tan \omega/\omega)^2}\,, \label{zw.1} \\
F_{\rm II}=F_2(\omega) &=& \frac{ \omega^2\left(1-\tan \omega/\omega
+\tan^2 \omega\right)}{3\,\left(1- \tan \omega/\omega\right)+
\tan^2 \omega}\,. \label{Fz.1}
\end{eqnarray}

\begin{figure}
\includegraphics[width=0.35\textwidth,clip=]{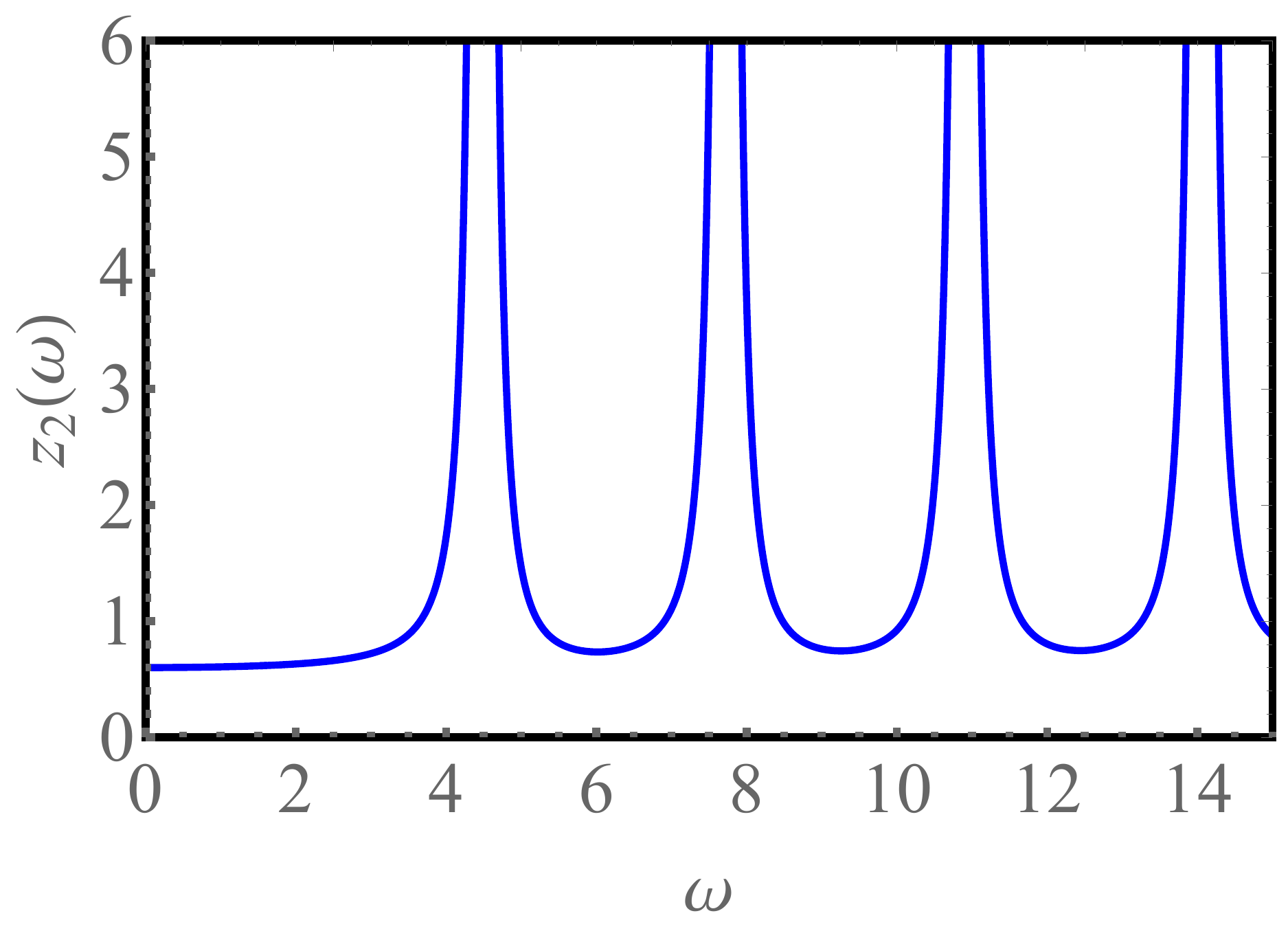}
\caption{The function $z_2(\omega)$ vs. $\omega$ from \eqref{zw.1}.
It has an infinite number of branches, but only the first branch corresponds to the optimal path. For this branch $z_2(\omega)$  diverges at
$\omega=\omega^*= 4.49341\dots$, which is the smallest positive root
of the equation $\tan \omega=\omega$.}
\label{fig:z2w_large}
\end{figure}
\begin{figure*}
   \begin{subfigure}[b]{0.31\textwidth}
        \includegraphics[width=\textwidth]{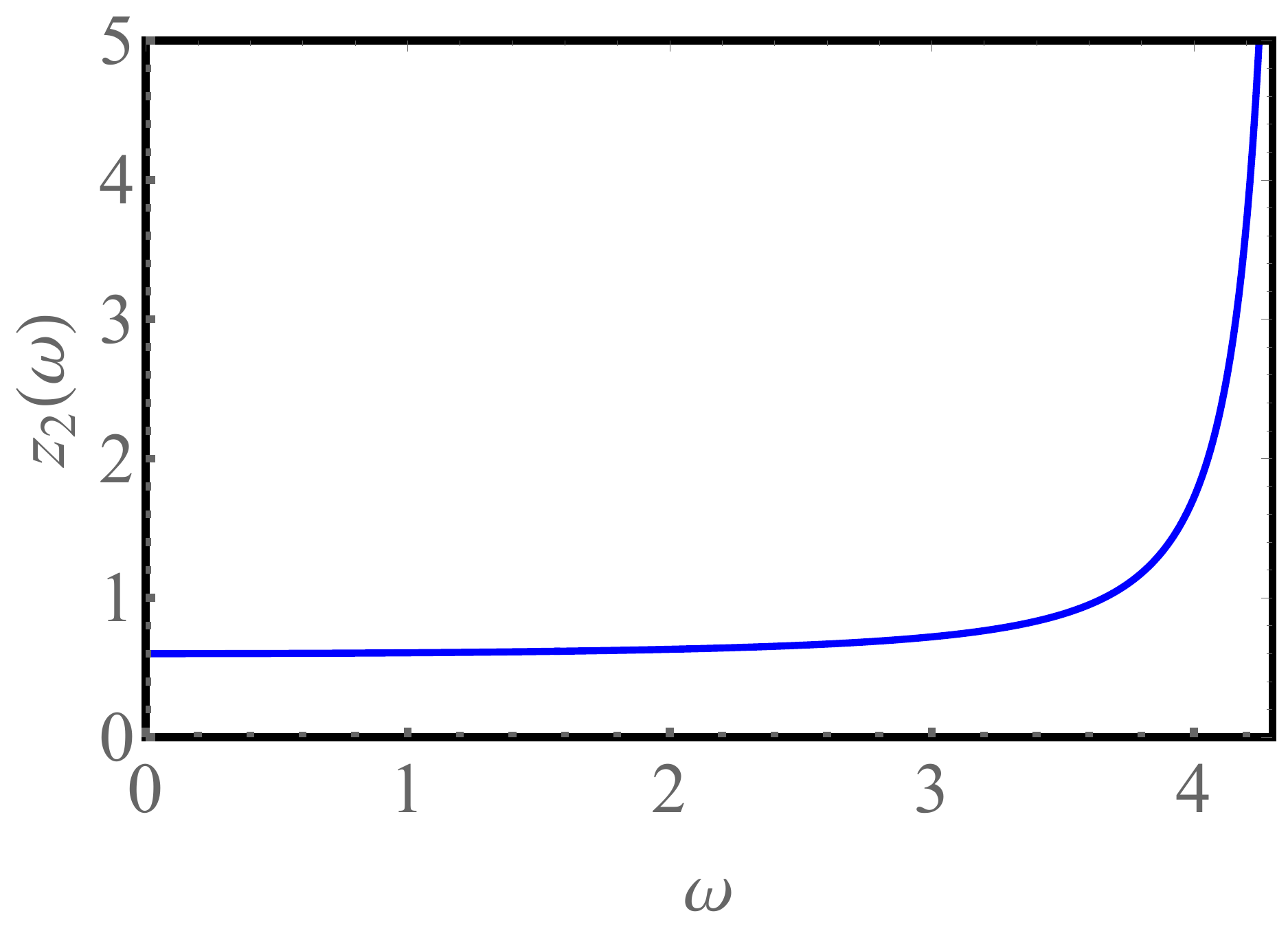}
        \caption{}
        \label{fig:zw2}
    \end{subfigure}
\quad
    \begin{subfigure}[b]{0.31\textwidth}
        \includegraphics[width=\textwidth]{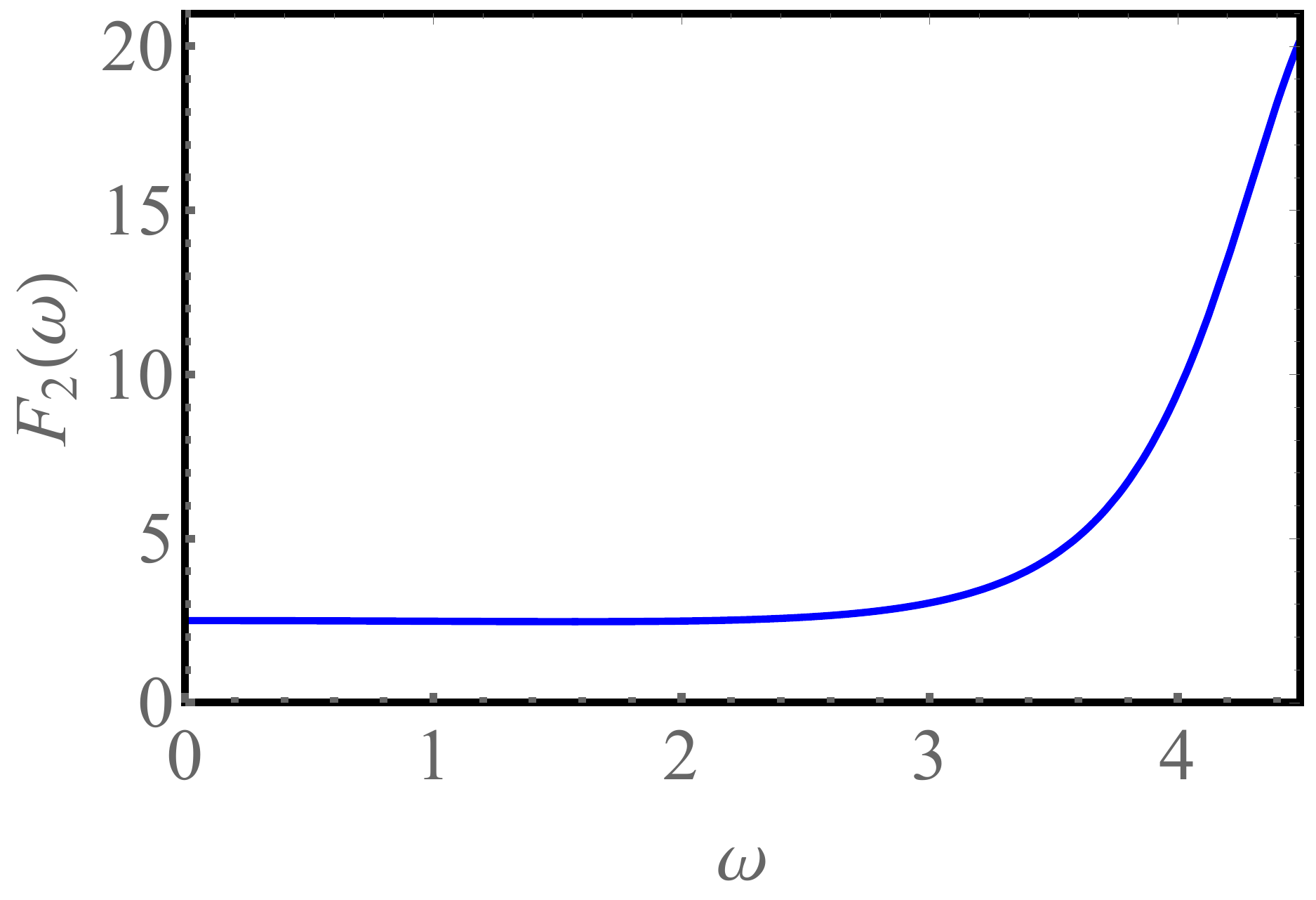}
        \caption{}
        \label{fig:Fw2}
    \end{subfigure}
\quad
    \begin{subfigure}[b]{0.31\textwidth}
        \includegraphics[width=\textwidth]{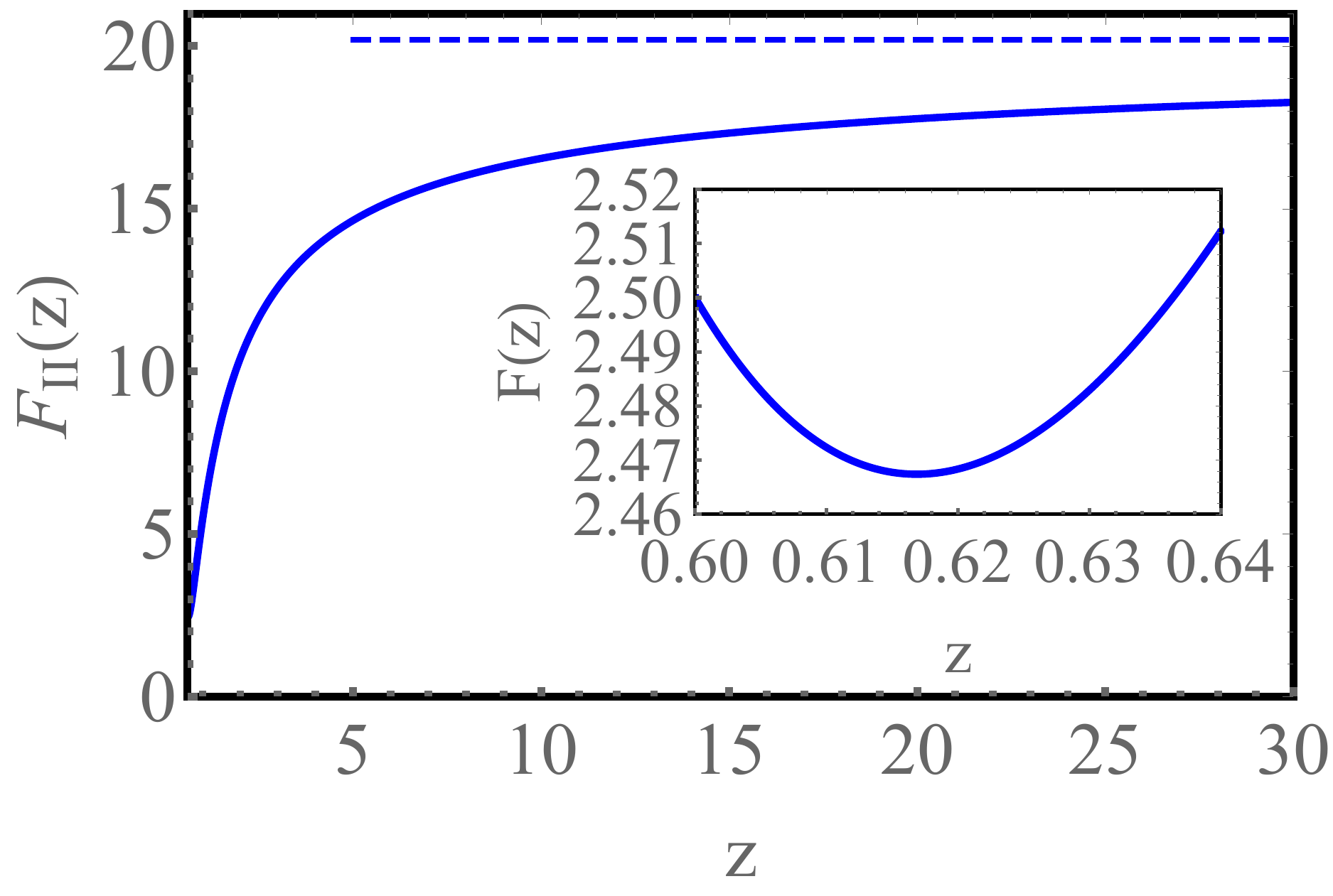}
        \caption{}
        \label{fig:Fz2}
    \end{subfigure}

\caption{(a) The function $z_{2}(\omega)$ vs. $\omega$ from \eqref{zw.1}. It increases monotonically
from $z_{2}(0)=3/5$ and diverges as $\omega$ approaches $\omega^*=4.49341$.
(b) The function $F_{2}(\omega)$ vs. $\omega$, as defined by \eqref{Fz.1}.
It starts from $F_{2}(0)=5/2$, slightly decreases [see the inset in (c)], and then grows monotonically
until $\omega= \omega^*=4.49341\dots$, reaching the value $F_{2}(\omega^*)= (\omega^*)^2= 20.1907\dots$.
(c) The function $F_{\rm II}(z)$ vs. $z$ , defined for $3/5\leq z<\infty$ and described in a parametric form by
Eqs.~\eqref{Fz.1} and \eqref{zw.1}. The horizontal dashed line is the large-$z$ asymptote
$F_{\rm II}(z\to \infty)=(\omega^*)^2= 20.1907\dots$.  The function $F_{\rm II}(z)$  reaches a minimum value $\pi^2/4$  at $z=\pi^2/16=0.61685\dots$, see the inset.}
\label{fig:branchII}
\end{figure*}
Note that the function $z_2(\omega)$ in \eqref{zw.1}
has multiple branches, see Fig.~\ref{fig:z2w_large}. As a result, for a fixed $z=z_2(\omega)$, there are multiple
solutions for $\omega$. The function, associated with the first branch in Fig.~\ref{fig:z2w_large},
diverges at  $\omega=\omega^*= 4.49341\dots$, which
is the smallest positive root of the equation $\tan \omega=\omega$, corresponding to a pole in Eq.~\eqref{zw.1}. As one can show,
the solutions, corresponding to the other branches, lead to larger values of the action $s= (1-X) F_2(\omega)$
for a fixed $X$, where $F_2(\omega)$ is given in \eqref{Fz.1}. Hence, for the optimal (least action)
path, we must select the first branch in Fig.~\ref{fig:z2w_large}, \textit{i.e.} restrict ourselves to
$0\le \omega\le \omega^*$. In this range of $\omega$ the function $z_2(\omega)$ is
monotonically increasing, see Fig.~\ref{fig:zw2}, with
the limiting behaviors
\begin{eqnarray}
z_2(\omega)\simeq \begin{cases}
& \frac{3}{5}+ \frac{\omega^2}{175}+ \ldots\,,
\quad \quad \omega\to 0, \\
\\
& \frac{1}{4\, (\omega^*-\omega)^2} + \ldots\,,  \quad\quad \omega\to \omega^* \,.
\end{cases}
\label{asymp_z2w}
\end{eqnarray}
The function $F_{2}(\omega)$ vs. $\omega$ is plotted in Fig.~\ref{fig:Fw2}.
It starts from $F_{2}(0)=5/2$, first decreases with increasing $\omega$, reaches
a minimum value $\pi^2/4$ at $\omega=\pi/2$ and then grows monotonically untill
$\omega=\omega^*$. It has the limiting behaviors
\begin{eqnarray}
F_2(\omega)\simeq \begin{cases}
& \frac{5}{2}-\frac{\omega^2}{42}+ \ldots\,,
\quad \quad \omega\to 0, \\
\\
& (\omega^*)^2 -4\, \omega^* (\omega^*-\omega)+\ldots \,,\quad\quad \omega\to \omega^*\,.
\end{cases}
\label{asymp_F2w}
\end{eqnarray}
The scaling function $F_{\rm II}(z)$ for $z\ge 3/5$, determined parametrically
by Eqs.~\eqref{Fz.1} and \eqref{zw.1}, is plotted in Fig. \ref{fig:Fz2}.
As $z\to \infty$, $F_{\rm II}(z)$ approaches the constant $(\omega^*)^2=20.1907\dots$.
The limiting behaviors of $F_{\rm II}(z)$ are given by
\begin{eqnarray}
F_{\rm II}(z)\simeq \begin{cases}
& \frac{5}{2} - \frac{175}{42}\,(z-\frac{3}{5})+ \ldots \,, \quad\quad z\to \frac{3}{5}\,,\\
\\
& (\omega^*)^2 - \frac{2\, \omega^*}{\sqrt{z}} +\ldots\,,  \quad\quad z\to \infty\,.
\end{cases}
\label{limits_FIIz}
\end{eqnarray}
We will refer to the solution for $z\ge 3/5$ as branch II. Here the optimal path $\phi(\tau)$ is
monotonic on the time interval $\tau\in [0,1]$ for $0\le \omega<\pi$, and non-monotonic
for $\pi<\omega< \omega^*$.

\subsubsection{Summary}

\begin{figure}
\includegraphics[width=0.32\textwidth,clip=]{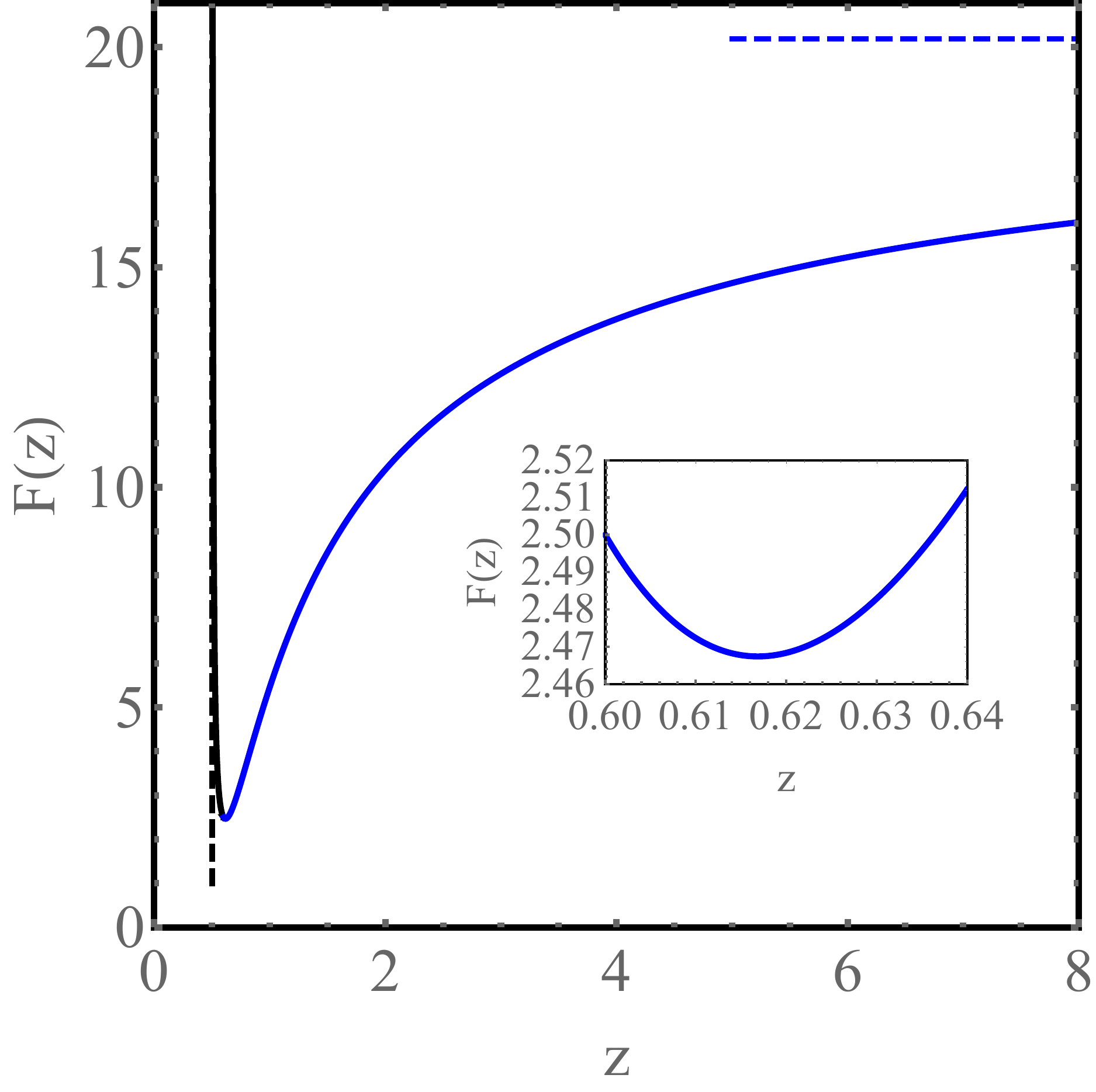}
\caption{The scaling function $F(z)$ in the full range $z\ge 1/2$, obtained
by combining $F(z)=F_{\rm I}(z)$ for $ 1/2\le z\le 3/5$ (branch I: the left one) and
$F(z)= F_{\rm II}(z)$ for $z\ge 3/5$ (branch II: the right one).  The horizontal dashed line is the large-$z$ asymptote
$F(z\to \infty)=(\omega^*)^2= 20.1907\dots$.  The vertical dashed line is the asymptote $z=1/2$.
The two branches join smoothly at $z=3/5$, where $F(z)=5/2$.  $F(z)$ reaches the minimum value $\pi^2/4$  at $z=\pi^2/16=0.61685\dots$, see the inset.
}
\label{fig:linearized}
\end{figure}

To summarize the results of this section, the joint probability $P(x,y,t)$ at short times, as predicted by the OFM at $1-X\ll 1$ and $|Y|\ll 1$, is given by Eq.~\eqref{rescaledaction} with the self-similar rate function
\begin{equation}
s(X,Y)= (1-X)\, F\left(\frac{1-X}{Y^2}\right)\,,
\label{s_scaling_linear}
\end{equation}
where
\begin{eqnarray}
F(z)=\begin{cases}
&\!\!\!F_{\rm I}(z),\; \frac{1}{2}\le z\le \frac{3}{5},\;\text{Eqs.~(\ref{zw_2.1}) and~(\ref{Fz_2.1})},
\\
\\
&\!\!\!F_{\rm II}(z),\; z\ge \frac{3}{5} , \;\text{Eqs.~(\ref{zw.1}) and~(\ref{Fz.1})}.
\end{cases}
\label{summaryF}
\end{eqnarray}
The branches I and II of the scaling function $F(z)$ join smoothly at $z=3/5$, where $F(z)=5/2$.  The asymptotic behaviors of $F(z)$ are the following:
\begin{eqnarray}
F(z)\simeq \begin{cases}
& \frac{1}{8\, (z-1/2)} + \ldots,
\quad \quad z\to 1/2, \\
\\
& (\omega^*)^2 - \frac{2\,\omega^*}{\sqrt{z}}\ldots \quad \quad  z\to \infty .
\end{cases}
\label{asympFz}
\end{eqnarray}
where $\omega^*=4.49341\dots$ is the smallest positive root of the equation $\tan \omega=\omega$. Figure~\ref{fig:linearized} shows a plot of
$F(z)$ over the full range $z\ge 1/2$.

\begin{figure}
\includegraphics[width=0.32\textwidth,clip=]{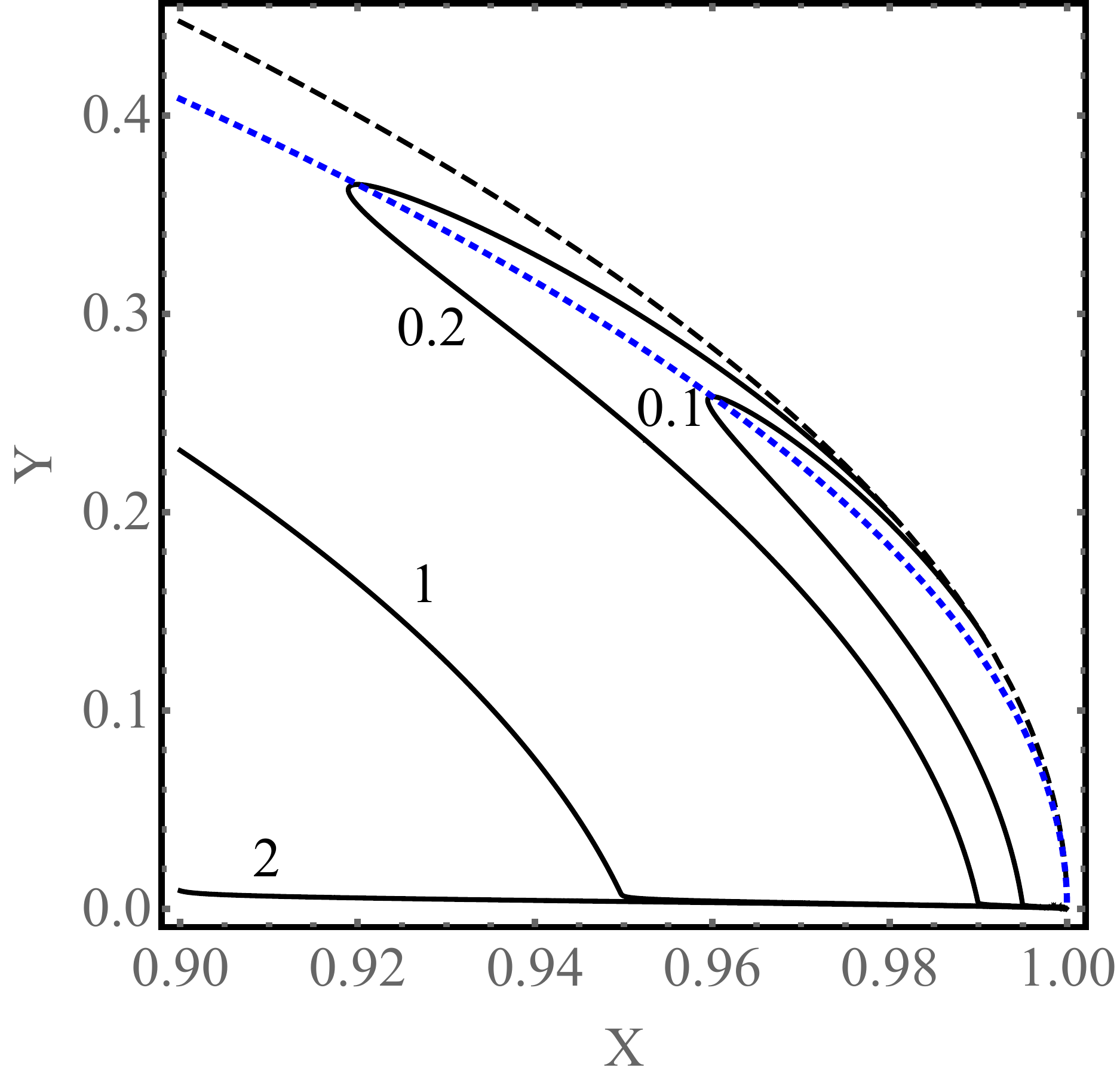}
\caption{The isolines $s=0.1$, $0.2$, $1$ and $2$ of the rate function $s(X,Y)$ near $X=1$ and $Y=0$, as described by
Eq.~\eqref{s_scaling_linear}. The dotted line $|Y|=\sqrt{(5/3) (1-X)}$ is the boundary between the two branches I and II.
The dashed line corresponds to the boundary of support of the distribution  $|Y|=\sqrt{2(1-X)}$, which is the leading-order approximation, close to  $X=1$ and $Y=0$, to the exact boundary of support $X^2+Y^2=1$.
The rate function in the region $Y<0$ (not shown) is symmetric, with respect to $Y=0$, to the rate function in the region $Y>0$.}
\label{fig:isolines}
\end{figure}
In its turn, Fig.~\ref{fig:isolines} shows a
contour plot of the rate function $s(X,Y)$,
described by Eq.~\eqref{action.3}. There is a single point $(X=1,\,Y=0)$ where $s=0$. This point corresponds to the trivial
(noiseless) optimal path $\phi(t)=0$. Then there are nested contours with increasing values of $s$ as one moves away from the point $(X=1,\,Y=0)$.  The OFM solution exists only for $z>1/2$, that is for $Y^2<2(1-X)$. Using the asymptotics of $F(z)$ in the first line
of \eqref{asympFz}, we see that, as $z\to 1/2$ from above, $s(X,Y)$ in Eq.\eqref{s_scaling_linear} diverges:
\begin{equation}
s(X,Y)\simeq \frac{1-X}{8(z-1/2)}\,,\quad \text{where}\quad z=\frac{1-X}{Y^2}\,.
\label{sxy_asymp.1}
\end{equation}
Consequently, the joint distribution $P(x,y,t)$ vanishes at $z=1/2$ extremely fast, via an
essential singularity:
\begin{eqnarray}
P(X,Y,t)&\sim &
\exp\left[- \frac{16(1-X)}{D t (z-1/2)}\right]\, .
\label{pxy_asymp.1}
\end{eqnarray}
Indeed, Fig.~\ref{fig:isolines} shows a very steep growth of $s$ as we approach the limiting parabola $Y^2=2(1-X)$. For $z\le 1/2$, that is at $Y^2>2(1-X)$, $P(x,y,t)$
is identically zero. All these features are in agreement with the exact analysis in the Introduction.
In particular, close to $X=1$ and $Y=0$,
the parabola $Y^2=2(1-X)$ is nothing but the leading-order approximation to the exact boundary of support of the joint distribution: the circle $X^2+Y^2=1$.

The self-similarity of the rate function~(\ref{s_scaling_linear}) close to the zero-noise point is a natural consequence of an invariance property of the governing equations~(\ref{s}), (\ref{X_cons}) and (\ref{Y_cons}) and the homogeneous boundary conditions (\ref{BCs_tau}). Under a stretching transformation  $\phi \to c\, \phi$, where $c=\text{const}$, $s$ becomes $c^2 s$, $1-X$ becomes $c^2 (1-X)$, and $Y$ becomes $c \,Y$. As a result, both $s/(1-X)$ and $(1-X)/Y^2$ remain invariant, leading to the self-similar structure of $s(X,Y)$.

Overall, the applicability conditions of our asymptotic~\eqref{rescaledaction} and (\ref{s_scaling_linear}) of the joint distribution $P(x,y,t)$ are
\begin{equation}\label{applicability}
D v_0 t^2\ll v_0t-x\ll v_0 t \quad \text{and}\quad |y|\ll v_0 t\,,
\end{equation}
where the strong inequality $D v_0 t^2\ll v_0t-x$ guarantees that the action $S$ from  Eq.~(\ref{action}) is much larger than unity.  Equations~\eqref{rescaledaction} and~(\ref{s_scaling_linear}) become asymptotically exact if we take the limits of $Dt\to 0$, $1-X\to 0$ and $Y\to 0$ while keeping the ratio $z=(1-X)/Y^2$ constant.

Finally, in Appendix A we show that the OFM prediction~(\ref{rescaledaction})  and~(\ref{s_scaling_linear}) for $P(x,y,t)$
in the vicinity of the point $(X=1, \,Y=0)$
perfectly agrees, at  $D v_0 t^2\ll v_0t-x$  with the ``near tail" of the expression for the
joint distribution of typical fluctuations,
derived by a different method in Ref.~\cite{Basu2018}.

\section{Summary and discussion}
\label{conclusion}

Even a single active Brownian particle (ABP) on the plane exhibits  unusual and non-intuitive properties.
Here we studied the position statistics of the ABP at short times, when the initial conditions are still important, and the activity aspects are the most pronounced. We employed the optimal fluctuation method (OFM), which is well suited for studying large deviations of the position coordinates $x$ and $y$ of the ABP from their expected values. Besides the large deviation functions themselves, the OFM predicts important additional observable quantities: the optimal paths of the particle, conditioned on reaching specified values of the coordinates.

At all times, the joint distribution $P(x,y,t)$ has a compact support, the boundary of which is a uniformly expanding circle
$x^2+y^2 =v_0^2 t^2$. The circumference of this circle includes the zero-noise point $(x=v_0t, \,y=0)$, where $P(x,y,t)$  has its maximum. The joint distribution vanishes in all other points of the circumference and outside of the circle.

Returning to short times: we showed that, close to the zero-noise point, the logarithm of $P(x,y,t)$ has an interesting self-similar structure, see Eqs.~\eqref{rescaledaction} and (\ref{s_scaling_linear}). This joint distribution matches smoothly with the tail of the joint distribution of typical fluctuations of $x$ and $y$, previously found by a different method in Ref.~\cite{Basu2018}. Determining the complete joint statistics of $x$ and $y$ inside the expanding circle is left for a future work. Here we determined the complete marginal $x$- and $y$-distributions, see Secs.~\ref{xdist} and~\ref{ydist}.  One important feature of the marginals is that they vanish extremely fast, via  essential singularities, at the edges of their support: at $x=-v_0 t$ for the $x$-marginal, and at $y=\pm v_0 t$ for the $y$-marginal. Since the corresponding $x$- and $y$- OFM actions diverge there, the essential singularities are not limited to short times and are expected to hold at all times. These large-deviation features are informative fingerprints of the short-time dynamics at long times, see also Ref. \cite{Basu2019}.

\section*{Acknowledgments} We are grateful to Naftali R. Smith for a useful comment. SNM
warmly acknowledges the hospitality of ICTS (Bangalore)
where this work was completed.
BM was supported by the Israel Science Foundation (Grant No. 807/16) and by a
Chateaubriand fellowship of the French Embassy in Israel. He is very grateful to the
LPTMC, Sorbonne Universit\'{e}, for hospitality.

\appendix

\section{Matching of the typical behavior of $P(x,y,t)$ with
its large deviation form obtained via OFM}
\label{Laplace_transform}

In Sec. \ref{joint_P} we employed the OFM to evaluate, up to a pre-exponential factor, the joint distribution $P(x,y,t)$
at short times near $X=1$ and $Y=0$. The results are given by Eqs.~(\ref{rescaledaction})  and~(\ref{s_scaling_linear}),
and they involve two branches of solution, I and II. Here we
show that these results match perfectly with the left tail of the expression for
$P(x,y,t)$,
which describes typical fluctuations of $x$ and $y$ and was obtained, by a different method, in  Ref.~\cite{Basu2018}.

To see this, we consider Eq. (8) of Ref.~\cite{Basu2018}, where the double
Laplace transform of the joint distribution in the region of typical fluctuations was computed at short times
near $x=v_0\,t$ and $y=0$. More precisely, it was shown in Ref.~\cite{Basu2018} that for $D t\ll 1$, and
setting $x-v_0t \sim t^2$ and $y\sim t^{3/2}$, the joint distribution in this typical regime takes
the scaling form as in \eqref{typical_scaling}.
[which of course is compatible with the exact scaling behavior~(\ref{Pexact})].
The double Laplace transform of the scaling function
${\tilde P}(a_1, a_2)$ as computed explicitly using path integral method~\cite{Basu2018}
\begin{equation} \int\int{\tilde P}(a_1,a_2)\, e^{-p_1a_1-p_2a_2}\, da_1\, da_2= \frac{\exp\left[p_2^2\, C(p_1)\right]}{\sqrt{\cosh\sqrt{2p_1}}}\, ,
\label{double_laplace.1}
\end{equation}
where
\begin{equation}
C(p_1)= \frac{1}{4\,p_1}\, \left(1- \frac{\tanh \sqrt{2p_1}}{\sqrt{2p_1}}\right)\, .
\label{Cp1.1}
\end{equation}
Inverting formally this Laplace transform, we get
\begin{equation} {\tilde P}(a_1,a_2)= \int
\frac{dp_1}{2\pi i}
\frac{e^{a_1 p_1}}{\sqrt{\cosh \sqrt{2p_1}}}
\int \frac{dp_2}{2\pi i} e^{p_2 a_2 +
p_2^2\, C(p_1)}\, ,
\label{invert.1}
\end{equation}
where the integrals are over the Bromwich contours in
the complex $p_1$ and the complex $p_2$ planes, respectively.
The integration over $p_2$ can be performed exactly
as it is just a Gaussian integral, leading to a single Laplace
integral in the complex $p_1$ plane,
\begin{equation} {\tilde P}(a_1,a_2)= \int \frac{dp_1}{2\pi i}
\frac{e^{a_1 p_1- a_2^2/{4\,C(p_1)}}}{\sqrt{4\,\pi\, C(p_1)\, \cosh \sqrt{2p_1}}}\,,
\label{invert.2}
\end{equation}
where $C(p_1)$ is given in \eqref{Cp1.1}. It is convenient to
rewrite this integral in a more suggestive form
\begin{equation}
{\tilde P}(a_1,a_2)= \int
\frac{dp_1}{2\pi i} \frac{e^{-\frac{a_1}{2}\, \mathcal{F}(p_1, z)}}{\sqrt{4\,\pi\, C(p_1)\, \cosh \sqrt{2p_1}}}\,,
\label{invert.3}
\end{equation}
where we set
\begin{equation}
\mathcal{F}(p_1,z)= \frac{1}{4\,z\, C(p_1)}- 2\, p_1 \quad {\rm and} \quad z=\frac{a_1}{2\, a_2^2}\,.
\label{saddle_action.1}
\end{equation}
So far Eqs. \eqref{invert.3} and \eqref{saddle_action.1}
are exact. It remains to perform the complex integral over $p_1$
in \eqref{invert.3}. The exact evaluation of this integral
seems difficult. However, for large $a_1$
and $a_2$, with the ratio $z=a_1/(2\, a_2^2)$ fixed,
we can evaluate the integral
by the saddle point method. The saddle point $p_1^*$ in the complex $p_1$ plane
is obtained from the saddle-point equation
\begin{equation}
\frac{\partial \mathcal{F}(p_1,z)}{\partial p_1}\Big|_{p_1=p_1^*}=0\, .
\label{saddle_point.1}
\end{equation}

By analyzing \eqref{saddle_point.1}, one can see that there is a
saddle point for all $z\ge 1/2$. For $1/2\le z\le 3/5$, the
saddle point occurs at a real positive $p_1$. Setting
$p_1=w^2/2$, we find, after some algebra, that the saddle point
equation \eqref{saddle_point.1}
exactly coincides with \eqref{zw_2.1} for the branch I of the OFM solution.
The associated saddle point action
$\mathcal{F}(w^2/2,z)$ yields
the OFM function $F_1(w)$  for branch I in
\eqref{Fz_2.1}.  Consequently, the tail of the joint distribution $P(x,y,t)$ of the typical fluctuations,
for large $a_1$ and $a_2$ but with
$z=a_1/(2\,a_2^2)$ fixed, behaves (up to a pre-exponential factor) as
\begin{equation}
P(x,y,t)\sim \exp\left[- \frac{a_1}{2}\, \mathcal{F}\left(\frac{w^2}{2}, z\right)\right]\,.
\label{Pjoint_tail.1}
\end{equation}
We recall that $a_1= (v_0\,t-x)/(v_0\, D \, t^2)$ and $a_2=y/(v_0\, \sqrt{2\,D}\, t^{3/2})$.
Therefore,
\begin{equation}
z= \frac{a_1}{2\, a_2^2}= \frac{ v_0\,t\,(v_0\,t-x)}{y^2}\, .
\label{z_typ}
\end{equation}
In terms of the rescaled positions $X$ and $Y$ we obtain
\begin{equation}
z= \frac{1-X}{Y^2}\,,
\label{z_typ.1}
\end{equation}
which coincides with the scaling ratio, defined in \eqref{ratioI.1}.
Similarly, the factor $a_1/2$ inside the exponent in \eqref{Pjoint_tail.1}
reads
\begin{equation}
\frac{a_1}{2}= \frac{1}{2 D t}\, (1-X) \, .
\label{a12.1}
\end{equation}
As a result, the tail of the
joint distribution $P(x,y,t)$, describing the typical fluctuations, matches exactly with
the OFM tail for $1/2\le z\le 3/5$.

Similarly, for $z\ge 3/5$, it turns out that the saddle point of \eqref{saddle_point.1}
is on the negative side of the real $p_1$ axis. Setting $p_1=-\omega^2/2$, and carrying out exactly the
same analysis as above, one can reconstruct the
OFM solution for branch II, \textit{i.e.}, for $\lambda_1\ge 0$.
Therefore,  the matching between
the asymptotics of the joint distributions $P(x,y,t)$ for the typical fluctuations and for the large deviations (the latter  described by the OFM)
occurs for all $z\ge 1/2$.

\section{Asymptotics of the marginal distributions from the joint distribution}
\label{marginalfromjoint}

Here we show how one can obtain the exponential asymptotic (\ref{neartailx}) of the near tail of $p_x(x,t)$, and the Gaussian asymptotic (\ref{Gaussy}) of $p_y(y,t)$, from the OFM expressions~(\ref{rescaledaction}) and~(\ref{s_scaling_linear}) for the joint distribution $P(x,y,t)$ near the point $(x=v_0t,\, y=0)$.

The marginal distribution $p_x(x,t)$ is given by the integral of $P(x,y,t)$  over $y$. At short times this integral can be evaluated by the saddle-point method. Working with our OFM expressions, which were derived up to pre-exponential factors, we should ignore any such factors in the calculation. We obtain, therefore,
\begin{equation}\label{px1}
 p_x(x,t) \sim \int\limits_{-\sqrt{2(1-X)}}^{\sqrt{2(1-X)}} dY\,\exp\left[-\frac{1-X}{2Dt}\,F\left(\frac{1-X}{Y^2}\right)\right].
\end{equation}
A change of the integration variable from $Y$ to $z=(1-X)/Y^2$ brings us to
\begin{equation}\label{px2}
 p_x(x,t) \sim \int_{0}^{\infty} dz\,\exp\left[-\frac{1-X}{2Dt}\,F(z)\right]\,
\end{equation}
where we ignored all pre-exponential factors, both inside and outside the integrals. The saddle point is at $z=\pi^2/16$, where $F(z)=F_{\rm{II}}(z)$ has its minimum value $\pi^2/4$, and we obtain
\begin{equation}\label{px3}
p_x(x,t) \sim \exp\left[-\frac{\pi^2(1-X)}{8Dt})\right] \,.
\end{equation}
Back to the variable $x$, this expression coincides with the near-tail asymptptic~(\ref{neartailx}).

\begin{figure}[h]
\includegraphics[width=0.30\textwidth,clip=]{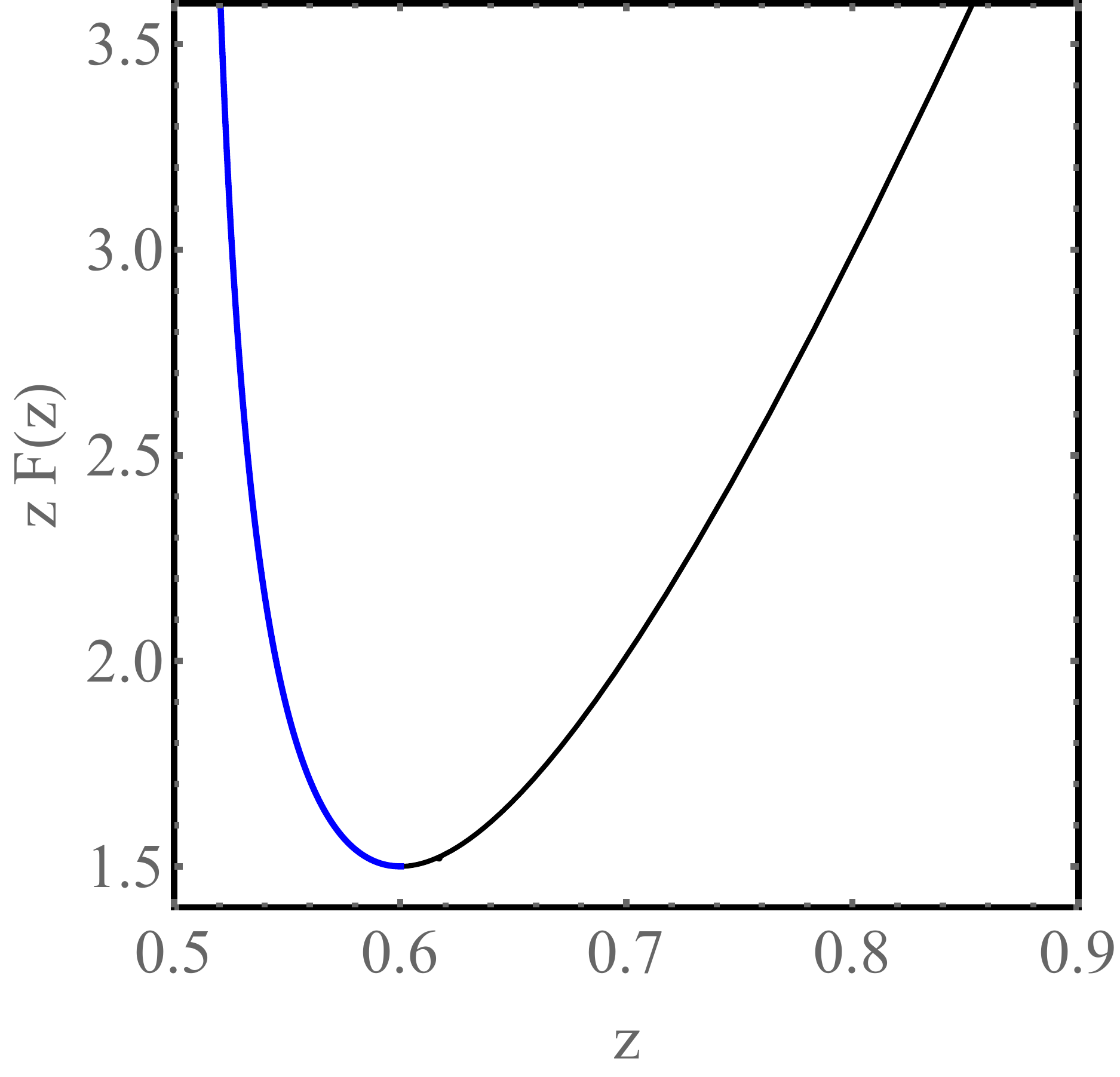}
\caption{A plot of the function $zF(z)$, where $F(z)$ is defined in (\ref{summaryF}).}
\label{zFz}
\end{figure}

The marginal distribution $p_y(y,t)$ is given by the integral of $P(x,y,t)$  over $x$, and we obtain
\begin{equation}\label{py1}
p_y(y,t) \sim \int\limits_{-1}^{1-Y^2/2} dX\,\exp\left[-\frac{1-X}{2Dt}\,F\left(\frac{1-X}{Y^2}\right)\right].
\end{equation}
The lower integration limit $-1$ is of course beyond the applicability domain of our $P(x,y,t)$, but this fact is inconsequential, because at short times the integral is dominated by a close vicinity of the upper integration limit. Going over from $X$ to $z$, we obtain
\begin{equation}\label{py2}
 p_y(y,t) \sim \int_{1/2}^{\infty} dz\,\exp\left[-\frac{Y^2}{2Dt}z F(z)\right]\,.
\end{equation}
Now the saddle point is the minimum point of the function $z F(z)$. A plot of the function $z F(z)$ vs. $z$ is shown in Fig. \ref{zFz}. The minimum value $3/2$ is achieved at $z=3/5$, that is at the matching point of the two branches $F_{\rm{I}}(z)$ and  $F_{\rm{II}}(z)$. As a result,
\begin{equation}\label{py3}
p_y(y,t) \sim \exp\left(-\frac{3Y^2}{4Dt}\right).
\end{equation}
This  Gaussian  can be properly normalized, leading to Eq.~(\ref{Gaussy}).


\begin{thebibliography}{99}
\bibitem{footnote1} There is also a monotone-decreasing solution, symmetric to this one with respect to the $x$-axis.
It has exactly the same action.
\end{thebibliography}
\end{document}